\documentclass[fleqn,10pt]{wlscirep}
\usepackage[utf8]{inputenc}
\usepackage[T1]{fontenc}
\usepackage{lineno}
\usepackage{soul,color}
\usepackage{textgreek}
\usepackage{booktabs}
\usepackage{graphicx}
\usepackage{tabularx}
\usepackage{multicol}
\usepackage{array}
\usepackage{makecell}

\soulregister{\cite}7
\soulregister{\ref}7

\title{QCDGE database, Quantum Chemistry Database with Ground- and Excited-state Properties of 450 Kilo Molecules}

\author[1,2]{Yifei Zhu}
\author[2]{Mengge Li}
\author[1,2]{Chao Xu}
\author[1,2,*]{Zhenggang Lan}

\affil[1]{SCNU Environmental Research Institute, Guangdong Provincial Key Laboratory of Chemical Pollution and Environmental Safety, MOE Key Laboratory of Environmental Theoretical Chemistry, South China Normal University, Guangzhou 510006, P. R. China.}
\affil[2]{School of Environment, South China Normal University, Guangzhou 510006, P. R. China.}

\affil[*]{corresponding author(s): Zhenggang Lan (zhenggang.lan@m.scnu.edu.cn)}

\begin{abstract}
    Due to rapid advancements in deep learning techniques, the demand for large-volume high-quality databases grows significantly in chemical research.
    We developed a quantum-chemistry database that includes 443,106 small organic molecules with sizes up to 10 heavy atoms including carbon (C), nitrogen (N), oxygen (O), and fluorine (F).
    Ground-state geometry optimizations and frequency calculations of all compounds were performed at the B3LYP/6-31G* level with the BJD3 dispersion correction, while the excited-state single-point calculations were conducted at the $\omega$B97X-D/6-31G* level.
    Totally  twenty seven molecular properties, such as geometric, thermodynamic, electronic and energetic properties, were gathered from these calculations.
    Meanwhile, we also established a comprehensive protocol for the construction of a high-volume quantum-chemistry database.
    Our QCDGE (Quantum Chemistry Database with Ground- and Excited-State Properties) database contains a substantial volume of data, exhibits high chemical diversity, and most importantly includes excited-state information.
    This database, along with its construction protocol, is expected to have a significant impact on the broad applications of machine learning studies across different fields of chemistry, especially in the area of excited-state research.

\end{abstract}

\begin{document}
\graphicspath{{Figures/}{Figures/scaffold}{Figures/scaffold_C}{Figures/fgs}}

\flushbottom
\maketitle

\thispagestyle{empty}

\section*{Background \& Summary}

In recent decades, 
the introduction of artificial intelligence (AI) and machine learning (ML) into chemistry research dramatically altered the paradigm of scientific discoveries.
With the development of computer science and technology, data played a more and more important role in several areas of chemical research.
Several chemical databases were created from the perspective of cheminformatics,
such as PubChem\cite{kim2023pubchem}, GDB\cite{fink2005virtual, blum970MillionDruglike2009, finkVirtualExplorationChemical2007, ruddigkeitEnumeration166Billion2012a}, ZINC\cite{sterling2015zinc,tingle2023zinc}, ChEMBL\cite{zdrazil2024chembl,davies2015chembl} and so on.\cite{pence2010chemspider,wishart2018drugbank}
These databases were widely used in various areas of chemistry,
which often serve as crucial data sources for
\textit{in silico} drug discovery, \cite{Cheng2014PubChem,Miller2002ChemicalDatabase,Bohacek1996StructureBased}
novel material development, \cite{Himanen2019DataDriven,Tripathi2020BigData,Cai2020MachineLearning,Zou2020OrganicLEDs,Salehi2019OLED,Zhao2020SolidState,Bruno2014Crystallographic}
\textit{etc}.

Recently, the development of molecular databases from first-principles quantum chemistry calculations attracted great attention.
The incorporation of these electronic-structure calculations largely improves the data consistency in the database,
eliminates inherent distribution errors, and provides molecular properties based on underlining atomic-level physical insights.
Therefore, this types of quantum-chemistry database shows the high transferable ability and the unified performance
across various applications.
Currently, available quantum-chemical databases can be roughly categorized into two main types: those that focuses the chemical and physical properties of different compounds with high diversities,
\cite{Montavon2013qm7b, kimEnergyRefinementAnalysis2019, ramakrishnanElectronicSpectraTDDFT2015, ramakrishnanQuantumChemistryStructures2014, rupp2012qm7,nakata2023pubchemqc, nakataPubChemQCPM6Data2020, nakataPubChemQCProjectLargeScale2017, chen2019alchemy,pereiraMachineLearningMethods2017,liangQMsymSymmetrizedQuantum2019,liangQMsymexUpdateQMsym2020}
and those that aim at exploring the non-equilibrium structures in the conformational space of specific molecules.

In recent years, 
the emergence of deep learning algorithms dramatically speed up the growth of the demand for a large-volume, high-quality database.  
In this new era of big data-driven chemical researches, two major limitations of existing molecular-property databases need to be addressed.

Firstly, the databases providing information on molecular excited states are very rare,\cite{ramakrishnanElectronicSpectraTDDFT2015, nakataPubChemQCProjectLargeScale2017, liangQMsymexUpdateQMsym2020,zouDeepLearningModel2023a} although the excited-state properties are immensely valuable in practical applications, ranging from photovoltaic devices,  organic light-emitting diodes, laser technologies and photobiological processes.
Therefore, there is an urgent need to develop the high-volume databases that contains high-quality excited-state data of molecules with large chemical diversities.

Secondly, with the development of the deep learning technologies,
the quality and quantity of data becomes a new bottleneck.
On the one hand, it is well known that very large data volumes can guarantee the correct interpolation ability and enhance the transferable abilities of ML models.
Although the simple combination of different databases seems to be a straightforward solution given their small overlap in chemical spaces, \cite{glavatskikhDatasetChemicalDiversity2019, isertQMugsQuantumMechanical2022}
this approach is not always recommended. \cite{kokkinos2017ubernet,zhangDPA2UniversalLarge2023}
The main issue is that the data from different databases do not match to each other due to their different resources.
In addition, many available databases still suffer from the lack of chemical diversity, and
this significantly deteriorates the performances of 
the deep ML models in chemical applications.\cite{glavatskikhDatasetChemicalDiversity2019}

Therefore, our objective is to build a quantum chemistry database that includes both ground- and excited-state properties.
This database must show massive-volume, data-consistency and high-diversity.
At the same time, the suitable protocol for the large database construction
must show a balance between effectiveness, efficiency, and accessibility.
To address this, we aim to develop such a comprehensive protocol
for general usages, which covers initial geometry selection, quantum chemical calculations, and data quality examinations, along with efficient accessing and retrieval processes.
This protocol ensures the robust and efficient construction of a large-volume database in chemical research.
We wish that the current work provides not only a valuable
large database but also a useful protocol to meet
the increasing tendency to treat large-volume data in future chemistry research.

In this work, we reported the QCDGE (Quantum Chemistry Database with Ground- and Excited-State Properties) database 
with high chemical diversity,
which totally includes 443,106 molecules
with up to ten heavy atoms within the carbon (C), nitrogen (N), oxygen (O) and fluorine (F) range.
These molecules are collected from the well-known QM9 \cite{ramakrishnanQuantumChemistryStructures2014}, PubChemQC\cite{nakataPubChemQCProjectLargeScale2017} and GDB-11 \cite{fink2005virtual, finkVirtualExplorationChemical2007} databases.
The ground-state geometry optimization and the frequency analysis were performed at the B3LYP/6-31G* level with the D3 version of Grimme's dispersion complemented by Becke-Johnson damping (BJD3), \cite{grimme2011effect}
while the excited-state single-point calculations including the first ten singlet and triplet transition states were performed at the $\omega$B97X-D/6-31G* level.
In total, 27 properties are extracted from these calculations, including ground-state energies, thermal properties, transition electric dipole moments, \textit{etc}.
We expect this database of small organic molecules to be useful in a wide range of applications in chemistry, especially for excited-state researches.

\section*{Methods}
\begin{figure}[ht]
    \centering
    \includegraphics[width=\linewidth]{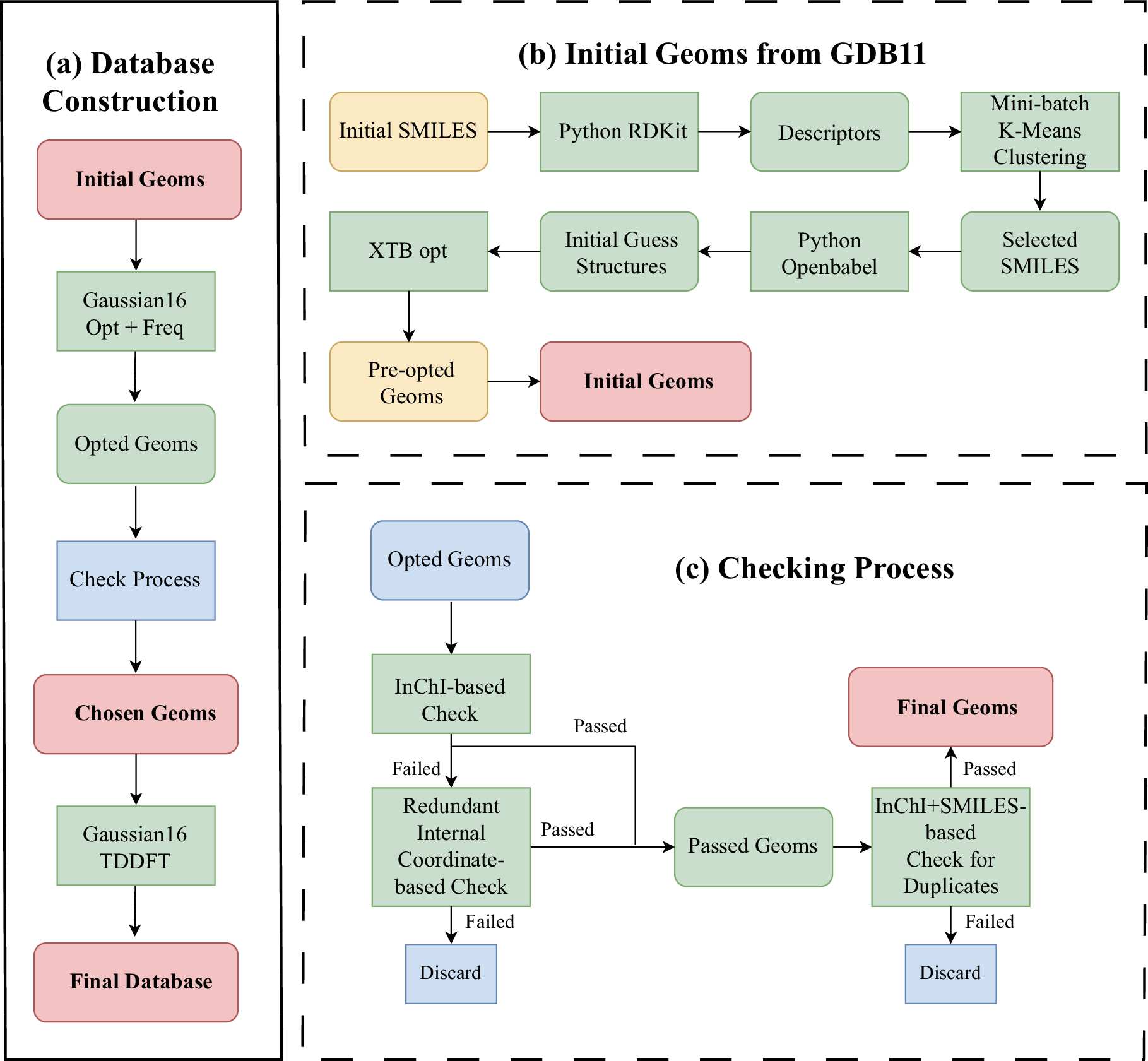}
    \caption{ 
        (a) The workflow employed in the data generation of the QCDGE database.
        (b) Initial geometry selection sourced from the GDB11 database.
        (c) Examination of optimization convergence and identification of duplicate geometries.
    }
    \label{fig:workfolow}
\end{figure}

In this work, we tried to construct the QCDGE database, which is a quantum chemistry database that contains ground-state and excited-state information of \textasciitilde450k molecules.
They are small organic molecules
with sizes up to ten heavy atoms in the range of C, O, N and F.
The database construction procedure is divided into four steps, as detailed in the following subsections.
The whole workflow is shown in Figure~\ref{fig:workfolow}(a).

\subsection*{Initial geometry collection}
Given our objective to build a database that reflects chemical diversity, it is imperative to ensure a balanced integration of data sources.
The GDB and PubChem databases commonly serve as molecular sources for the construction of quantum chemistry databases.
The GDB series databases are constructed through molecular combinatorial enumerations, according to the criteria of chemical stability and synthetic feasibility.
In contrast, PubChem data are built from hundreds of data sources, including government agencies, chemical vendors, journal publishers, and others,
which offers a broad collection of molecular information.
It was observed that the duplication of data between these two databases may not be significant, \cite{glavatskikhDatasetChemicalDiversity2019, isertQMugsQuantumMechanical2022}
which allows us to integrate molecules from these two databases to define a new one.
Therefore, we believe that the GDB and PubChem databases stand out as original data sources with the optimal balance between chemical diversity and reliability.

In the first stage, our aim was to collect molecules with a size of up to 9 heavy atoms.
To save computational resources, we first consider quantum chemistry datasets derived from GDB and PubChem, in which the three-dimensional optimized molecular structures were already given.
Specifically, we took molecules from two primary databases: QM9 and PubChemQC, which are derived from GDB-17 (also built from GDB) and PubChem, respectively.
The QM9 dataset contains 132,177 molecules
with sizes up to nine heavy atoms in the C, O, N and F range.
Because of its significance and reliability, it is generally considered one of the golden standard databases in the field of ML chemistry.
Therefore, we chose compounds from the QM9 dataset according to the above selection rule.
At the same time, we selected 122,785 molecules from PubChemQC, by using the same selection rule of QM9, in terms of the same limitation on heavy atoms.

In the second step, we broadened our selection criteria to include molecules with up to ten heavy atoms, in order to cover a wider chemical space.
As the PubChemQC database contains many large compounds, we simply extracted a subset of 105,085 molecules that meet this new standard.
As contrast, no such molecules are found in the QM9 dataset.
Therefore, we selected additional molecules from GDB-11,
a database generated from the original GDB.
It is necessary to mention that many databases were derived from GDB, such as GDB-11, GDB-17, QM9, etc.
In principle, the GDB-17 database might be a better choice
since the QM9 dataset is derived from it.
However, in practice, the large size of the GDB-17 database brings numerical challenges.
Because both GDB-11 and GDB-17 are generated on the basis of the same algorithm, we chose the smaller GDB-11 database here.

Here GDB-11 still contains over 3,000,000 molecules characterized by 10 heavy atoms (C, N, O and F elements).
The direct inclusion of such large amount of molecules would disrupt the balance of data distributions.
To manage this, we used the mini-batch K-Means clustering algorithm \cite{sculley2010web} to divide these molecules into 10,000 clusters.
We then randomly selected molecules from these clusters, ensuring that the number of molecules chosen from each cluster was proportional to ratio between the cluster size and the total number of molecules.
In this way, we selected 134,681 molecules,
achieving a balanced representation across the chemical space.

Given that GDB-11 solely offers SMILES representations for its molecules, we needed to perform the initial geometry optimization. For this purpose, Cartesian coordinates of these molecules were generated using the in-house Python interface to Open Babel (version 2.8.1). \cite{o2008pybel, o2011open}
Subsequently, these geometries were optimized using the semi-empirical method GFN2-xTB \cite{bannwarth2019gfn2} in the xtb program\cite{bannwarth2021extended}.
The whole process of collecting initial geometries from GDB-11 is shown in Figure~\ref{fig:workfolow}(b).
For the in-depth information
ranging from molecular descriptors, clustering methods, to the data selection strategies, please refer to the Supplementary Information (SI).

Finally, we collected 494,701 pre-optimized molecular structures with balanced data sources, in which all small organic molecules at most contain ten heavy atoms of C, N, O, and F.

\subsection*{Ground-state calculations}
Once the initial geometries were collected, we moved to the step of the quantum chemistry calculations, which is the most time-consuming one.
Here, ground-state geometry optimizations and frequency analyses were conducted for all molecules.
These calculations were performed at the DFT level using the B3LYP  functional and the 6-31 G(d) basis set, which employ the Gaussian 16 package (B.01 version). \cite{g16}
To enhance accuracy, we incorporated the BJD3 dispersion corrections.
Among all optimization jobs, about 0.3\% (1,534) calculations failed to converge.
This indicates that most initial structures are reasonable.

\subsection*{Optimized geometry check}
In order to check the optimized geometries and to streamline the database by removing duplicated compounds, we proposed an examination process as shown in Figure~\ref{fig:workfolow}.(c).

The first goal of this step is to confirm the consistency between optimized geometries and their initial counterparts.
We generate InChI (IUPAC International Chemical Identifier) strings \cite{heller2015inchi} of initial and optimized geometries
with the Python scripts interfaced with Open Babel (version 2.8.1).
In most situations, the initial and optimized geometries give consistent InChI representations,
indicating the reliability of the optimization tasks.
Occasionally, some pairs show obvious discrepancies.
This may refer to situations where the optimized geometry and the initial geometry are significantly different, implying that the optimization may not obtain a consistent result. 
However, such discrepancies may simply be
due to the fact that the definition of InChI codes
is too rigorous, and this very tight rule largely exaggerates stereoisomeric differences, even for minor ones.
Therefore,  when initial and optimized molecular structures show different InChI representations, additional examinations of geometrical details should be performed to avoid misjudgments.
In practice, the redundant internal coordinates \cite{pulay1992geometry,peng1996using} 
of the initial and optimized geometries were extracted using the Gaussian 16 software.
The direct comparison of them gave us solid answers to
address whether the optimization task brings significant changes in molecular structures.
When the optimization task gives the consistent structure with respect to the initial one, the molecule was retained in the database.

The second target is to eliminate duplicate geometries from our database.
To achieve this, we simultaneously compare the structures using both SMILES and InChI strings generated via Open Babel.
Given that these two representations highlight different aspects of the molecular geometries, 
it is enough to use them to classify duplicated molecules.
In addition, this detection methodology shows the effective balance between computational accuracy and efficiency.

After removing 51,595 geometries that failed in optimization 
(1,534) and belonged to duplicated ones (50,061), finally the refined database contains 443,106 geometries.
The relatively low proportion of duplicates further confirms
that two original databases cover different areas of the chemical space.

\subsection*{Excited-state calculations}
All excited-state calculations were performed using the TDDFT method with the $\omega$B97X-D functional and the 6-31 G(d) basis set.
The reason to choose the $\omega$B97X-D functional is mainly due to the fact that it gives reasonable descriptions of the charge transfer states, while the employment of the B3LYP level
here may significantly underestimate the excitation energies of the charge transfer states.
In the TDDFT calculations, the first ten singlet and triplet excited states were included.
These calculations were carried out with Gaussian 16 software, employing the molecular geometries optimized at the B3LYP/6-31G(d)/BJD3 level.

\section*{Data Records}

All data, including optimized molecular structures and important molecular properties, were extracted from the results of the quantum chemistry calculations.
They are organized in a standard manner,
which are accessible either in the figshare repository and or on the website of this data-driven excited-state information project \cite{langroup}.
To ensure the integrity of all data in the further applications, we also provide the corresponding 512-bit cryptographic hash 
generated by the Secure Hash Algorithm 512 (SHA-512) for verification.

In the uploaded files, the $final\_all.csv$ summarizes the basic information of all molecules, such as their features
(SMILES and InChI strings),
the number of the heavy atoms, the number of ring moieties and so on.
Within this file, the string in the first column serves as the unique identifier for each molecule.
All data obtained from ground- and excited-state quantum chemistry calculations are saved 
in a binary file with the compressed version of the Hierarchical Data Format version 5 (HDF5) \cite{osti_1631295} format.
The HDF5 format is specifically designed to handle large volumes of numerical data, offering more efficient disk space utilization compared to other file formats such as text, JSON, and YAML.
It supports reading data in chunks, enhancing efficiency in complex data analysis.
The compressed version of HDF5 further reduces the record space significantly.
In the current compressed version of the HDF5 file, the information of each molecule is organized as a dataset named after its identifier.
Several versions of the SMILES and InChI strings are assigned as attributes for the dataset.
Within each molecular dataset, 14 ground-state and 13 excited-state properties were recorded, as described in Table \ref{tab:props}.

\section*{Technical Validation}

The chemical diversity and data quality of the current QCDGE database were examined in different ways, see below.
In this section, all analyzes were performed based on the Python interface of RDKit (version 2023.3.1).

Here, given that the current database does not contain duplicated structures, we simply
divided all data into two subsets, namely \textit{data\_A} and \textit{data\_B} according to their original resources, 
the GDB series databases and the PubChemQC database, respectively.
These labels are used mainly for better illustrations
in the following discussion.
The differences in the chemical spaces of \textit{data\_A} and \textit{data\_B} were discussed in the SI.
We wish to emphasize that the conclusions drawn from the forthcoming analysis of \textit{data\_A} and \textit{data\_B} can not be generalized to describe the properties of
the GDB and PubChemQC databases themselves.

\subsection*{Element composition}
Fifteen elemental compositions were identified according to different combinations of four heavy atoms (C, N, O and F).
The numbers of compounds in several leading composition groups are given in Table~\ref{tab:composition}.
Among them, the group composed of molecules containing three heavy elements (C, N, and O) at the same time is the largest one, accounting for slightly less than the half of the total.
The molecules including both C and O atoms together, as well as C and N, define the second and and third largest groups, their total contributions accounting for about 3/4 of the total.

Same analysis were conducted on two molecular subsets, \textit{i.e.}, \textit{data\_A} and \textit{data\_B},
as shown in the Table S1 and S2.
The results clearly demonstrate that the data from two different resources in fact do not overlap with each other in the chemical space.
Interestingly, \textit{data\_A} has a high proportion of molecules that contain other heavy atoms except C, whereas \textit{data\_B} exhibits a greater diversity of carbon skeletons.
However, molecules containing solely N atoms, solely O atoms, those containing both O and F, both N and F, and those containing N, O and F simultaneously, only appear in \textit{data\_B}.

\subsection*{Topology}
\begin{figure}[ht]
    \centering
    \includegraphics[width=\linewidth]{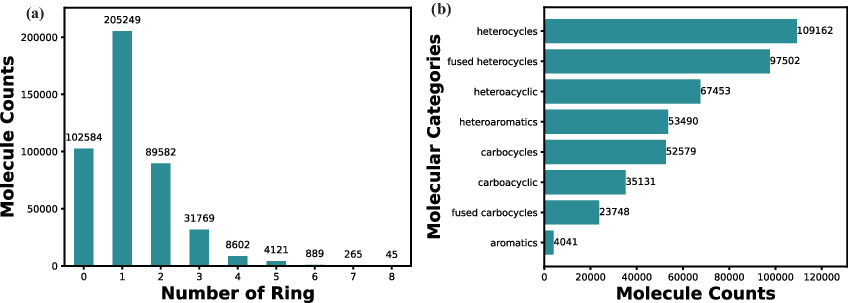}
    \caption{
        (a) Distribution of molecules by the number of rings.
        The histogram illustrates the counts of molecules categorized by their corresponding ring numbers.
        (b) Diversity of molecular categories.
        The horizontal axis quantifies the number of molecules examined, while the vertical axis lists several types of molecules discovered. Each bar represents the frequency of a particular molecular type.
    }
    \label{fig:fig2}
\end{figure}

As shown in Figure~\ref{fig:fig2} (a), over 3/4 of the molecular structures contain ring moieties, and nearly the half of all molecules only include a single ring from a topological perspective.
Among them, the proportions of acyclic molecules in \textit{data\_A} and \textit{data\_B} are approximately \textasciitilde15\% and \textasciitilde35\%, respectively (Figure S1).
On average, molecules in \textit{data\_A} possess 1.51 rings, whereas in \textit{data\_B}, the average is 0.81 rings.
This finding prompts us to make a more comprehensive investigation of the existence of various ring units,
as the ring moieties, particularly the aromatic rings, 
play important roles to determine excited-state properties.

\subsection*{Compound type}
According to the descending order of the number of molecules,
all composition groups were sorted as follows:
heterocycles (24.6\%), fused heterocycles (22.1\%), heteroacyclic (15.3\%), heteroaromatics (11. 9\%),  carbocycles
(11. 9\%), carboacyclic compounds (7.9\%), fused carbocycles (5.4\%), and aromatics with carbon rings (0.9\%).
This distribution is consistent with the chemical intuitive notion, as the introduction of heteroatoms should largely extend the chemical space with respect to the situations with only carbon atoms.

Significant differences appear in the distributions of compound types in \textit{data\_A} and \textit{data\_B}, as illustrated in Figure S2.
This divergence could be attributed to the following reasons.
More ring structures appears in \textit{data\_A}, and thus the fused heterocycles are popular.
In contrast, \textit{data\_B} contains more acyclic compounds, including both heteroacyclic and carboacyclic ones.
As the consequence, \textit{data\_A} features a greater complexity of ring moieties, whereas \textit{data\_B} is characterized by a relative abundance of linear or non-ring structures.

\subsection*{Scaffold analysis}
Using the RDKit toolkit, the Murcko scaffold analysis was conducted to explore the diversity of molecular backbone in the QCDGE database.
Aside from acyclic molecules (\textasciitilde23.2\%),
totally 59,898 distinct scaffolds were identified among the remaining molecules.
Among all identified scaffolds, the most dominant one is three-membered carbon ring (C1CC1) moieties, as shown in Table~\ref{tab:all_atom}.
This may be attributed to the following fact. 
As our selection rule only chose molecules limited to ten heavy atoms, both small and large molecules may easily include stable three-membered rings. 

To facilitate a more comprehensive analysis, we can also make Murcko scaffolds generic as illustrated in Table~\ref{tab:all_C}, by converting all types of atom to carbon and treating all bonds as single bonds.
In such analysis, 3,258 Murcko scaffolds were identified,
while five-membered rings predominated.

The scaffold analysis were also carried out in \textit{data\_A} and \textit{data\_B}, and results are detailed in Tables S3 to S6.
In the standard scaffold analysis, 46,234 scaffolds were identified in \textit{data\_A} and 17,290 in \textit{data\_B}.
In contrast, the generic scaffold analysis yielded 2,161 and 1,934 scaffolds for \textit{data\_A} and \textit{data\_B}, respectively.
This observation suggests that \textit{data\_A} show higher chemical diversity in the ring part than \textit{data\_B},
consistent with their individual features.

\subsection*{Functional group analysis}
The diversity of functional groups was explored using the Ertl algorithm \cite{ertlAlgorithmIdentifyFunctional2017a, schaubdevelopment}, achieved with the RDKit (version 2023.3.1) toolkit.
Initially, the original RDKit version only recognizes a limited range of generic functional groups composed of C, N, O, and F.
To enhance the analysis ability, we expanded its functionality to identify 109 functional groups (as shown in Figure S3), according to  their definitions in Checkmol software. \cite{haider2010functionality}
The current in-house expansion mainly improves the analysis protocol in two ways, (i) making the distinction of substituents such as dialkylether and alkylarylether; (ii) including some larger functional groups such as hemiaminal.

Across the dataset, 102 functional groups were detected and the number of functional groups on average is 2.4 per molecule.
Top twenty functional groups are shown in Table~\ref{tab:fgs}.
Importantly, the absence of certain functional groups in our database does not suggest a lack of chemical diversity,
while it may be attributed to the limitation on the number of atoms due to our selection rule.

The analysis of functional groups in molecules from \textit{data\_A} and \textit{data\_B} showed different distributions, as detailed in Tables S7 and S8.
102 and 98 types of functional groups were identified within \textit{data\_A} and \textit{data\_B}, respectively.
The similar numbers here suggest that both datasets display very high degrees of chemical diversity.
However, this does not imply that the distribution of chemicals in two datasets is similar.
For the same functional group, it is clear that
its proportion is different in two subgroups.
These results highlight the differences in chemical diversity between two subgroups, and further confirm the importance of merging two data sources.

\subsection*{Excitation energy}
\begin{figure}[ht]
    \centering
    \includegraphics[width=\linewidth]{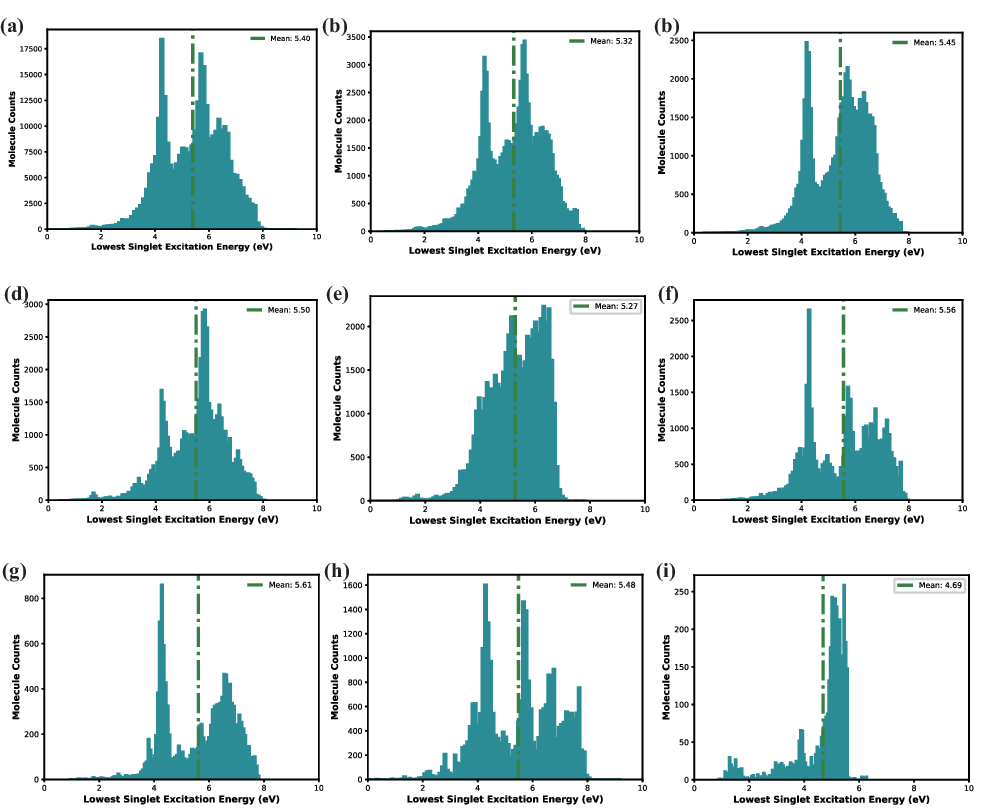}
    \caption{ The distribution of the lowest singlet state excitation energy across (a) all selected molecules with double or triple bonds, and (b) heterocycles, (c) fused heterocycles, (d) heteroacyclic, (e) heteroaromatics,  (f) carbocycles, (g) carboacyclic compounds, (h) fused carbocycles, and (i) aromatics with carbon rings.
    }
    \label{fig:all_es}
\end{figure}

We tried to analyze the excited state properties saved in our QCDGE database.
Considering that functional groups containing double or triple bonds are typically responsible for the photoexcitation to the low-lying excited states of molecular systems, we mainly focus on molecules containing such bonds.
In the QCDGE database, 346,312 molecules,
representing over 78\% of total molecules,
contain double or triple bonds.
The excitation energies of the lowest singlet states of them are shown in Figure~\ref{fig:all_es} (a).
The vast majority of these molecules display the lowest singlet state excitation energies distributed between 2 and 8 eV.
Since the excitation energy is closely relevant to the type of compound, 
the corresponding distributions are given in Figure~\ref{fig:all_es}.
Among all compound types, aromatic compounds have the lowest average singlet state excitation energies, while carboacyclic compounds have the highest values.
This observation is highly consistent with chemical intuition.
Additionally, the distribution of the lowest singlet-state excitation energies across all molecules in the QCDGE database is shown in Figure S4.

\section*{Usage Notes}
We offer a Python script named \textit{extract\_data.py}, designed to extract relevant data from HDF5 files.
This script allows for extracting molecular properties from the QCDGE database,
in which many options are supported as well.
It can process the full list of all molecules in the database, a predefined list of molecules, or a chosen set of molecules filtered by the number of heavy atoms and their elemental compositions.
It is also possible to import the $extractData()$ class from this script, providing the seamless integration with other Python codes.
All script and data files are available in the figshare repository and the project website \cite{langroup}.

\section*{Code availability}
All research was supported by the Python programming language (version 3.8.5) \cite{python385}, while several important Python libraries and their respective versions are outlined below.
Open Babel (version 2.8.1) and RDKit (version 2023.3.1) Python libraries were used to generate cheminformatic representations and to perform analysis.
The management of HDF5 files was facilitated by h5py (version 2.10.0)\cite{andrew_collette_2019_3401726}, while pandas (version 1.1.3)\cite{jeff_reback_2020_4067057,mckinneyprocscipy2010} were used to implement CSV files and perform relevant data analysis.
All related scripts are also available on GitHub \cite{codegithub}.
All scripts fall into three categories: calculation, check, analysis, while a separate Python script \textit{extract\_data.py} is also given to extract data information.

\bibliography{main}

\section*{Acknowledgements}
This work was supported by NSFC projects (Nos.22333003, 22361132528 and 21933011) and the Opening Project of Key Laboratory of Optoelectronic Chemical Materials and Devices of Ministry of Education, Jianghan University (JDGD202216).

\section*{Author contributions statement}
Conceptualization: Z.L. Data Curation: Y.Z. and M.L. Formal Analyses: Y.Z. and Z.L. Funding Acquisition: Z.L. Investigation: Y.Z., M.L. and Z.L. Methodology: Y.Z. Project Administration: Z.L. Resources: C.X. and Z.L. Software: Y.Z. and C.X. Supervision: C.X. and Z.L. Validation: Y.Z. and M.L. Visualization: Y.Z. and M.L. Writing - Original Draft Preparation: Y.Z. Writing - Review \& Editing: Y.Z., M.L., C.X. and Z.L.

\section*{Competing interests}
The authors declare no competing interests.

\section*{Figures \& Tables}

\begin{table}[ht]
\centering
\begin{tabularx}{\textwidth}{|X|>{\centering\arraybackslash}m{3cm}|>{\centering\arraybackslash}m{3cm}|>{\centering\arraybackslash}m{3cm}|>{\centering\arraybackslash}m{4.5cm}|}
\hline
Database &  Chemical space & Number of selected molecules up to 9 heavy atoms & Number of selected molecules with 10 heavy atoms & Molecular structure data \\
\hline
PubChemQC & PubChem &  122,758 & 105,085 & Cartesian coordinates\\
\hline
QM9 & GDB series & 132,177 & 0 & Cartesian coordinates \\
\hline
GDB-11 & GDB series & 0 & 134,681 & SMILES strings \\
\hline
\end{tabularx}
\caption{\label{tab:initial}494,701 initial data sources.}
\end{table}

\begin{table}[ht]
    \centering
    \begin{tabular}{llll}
    \toprule
    No. & Source & Key in HDF5 & Description \\
    \midrule
    1&  GS &    labels           & Atomic labels. \\
    2&  GS &    coords           & Optimized Cartesian coordinates.\\
    3&  GS &    Etot             &  Total energy.\\
    4&  GS &    e\_homo\_lumo    &  HOMO and LUMO Energies.\\
    5&  GS &    polarizability   &  Isotropic polarizability.\\
    6&  GS &    dipole           & Dipole moment. \\
    7&  GS &    quadrupole       &  Quadrupole moment.\\
    8&  GS &    zpve             & Zero-point vibrational energy. \\
    9&  GS &    rot\_constants   & Rotational constant. \\
    10&  GS &    elec\_spatial\_ext & Electronic spatial extent. \\
    11&  GS &    thermal          & Thermal properties at 298.15 K.\\
    12&  GS &    freqs            & Harmonic vibrational frequencies. \\
    13&  GS &    mulliken         & Mulliken charges. \\
    14&  GS &    cv               & Heat capacity at 298.15 K. \\
    \midrule
    1& ES &     Etot                                   & Ground-state energy. \\
    2& ES &     e\_homo\_lumo                            & HOMO and LUMO Energies \\
    3& ES &     dipole                                 & Dipole moment. \\
    4& ES &     quadrupole                             & Quadrupole moment. \\
    5& ES &     rot\_constants                          & Rotational constant. \\
    6& ES &     elec\_spatial\_ext                       & Electronic spatial extent  \\
    7& ES &     mulliken                               &Mulliken charges.   \\
    8& ES &     transition\_electric\_DM                 & Transition electric dipole moments.  \\
    9& ES &     transition\_velocity\_DM                 & Transition velocity dipole moments. \\
    10& ES &     transition\_magnetic\_DM                 & Transition magnetic dipole moments. \\
    11& ES &     transition\_velocity\_QM & Transition velocity quadrupole moments. \\
    12& ES &     OrbNum\_HomoLumo                        & Orbital numbers of  HOMO and LUMO. \\
    13& ES &     Info\_of\_AllExcitedStates               & Electronic characters of 10 singlet and 10 triplet excited states.  \\
    \bottomrule
    \end{tabular}
    \caption{\label{tab:props}
    The fundamental and calculated information extracted from both ground-state and excited-state quantum chemistry calculations.
    Due to the utilization of different functionals in ground-state and excited-state calculations, some properties are extracted in both scenarios.
    In the \textit{Source} column, \textit{GS} and \textit{ES} indicates the property obtained from the calculation of the ground and excited state, respectively.
    }
\end{table}

\begin{table}[ht]
    \centering
    \begin{tabular}{lrrrrrrrrrrrr}
    \toprule
    \multicolumn{12}{c}{Element Composition} \\
    \cline{2-11}
    \rule{0pt}{2.5ex}
    Number of\\ heavy atoms &\textbf{C}&\textbf{N}&\textbf{NO}&\textbf{CN}&\textbf{CO}&\textbf{CF}&\textbf{CNO}&\textbf{CNF}&\textbf{COF}&\textbf{CNOF}&\textbf{Total} \\
    \midrule
    \textbf{2}&4&3&2&3&3&1&0&0&0&0&\textbf{16}\\
    \textbf{3}&7&4&9&21&17&4&10&5&2&0&\textbf{79}\\
    \textbf{4}&32&7&18&114&70&19&108&23&20&9&\textbf{420}\\
    \textbf{5}&92&6&20&418&283&58&501&60&82&39&\textbf{1559}\\
    \textbf{6}&279&5&20&1325&1021&149&1878&135&239&89&\textbf{5140}\\
    \textbf{7}&683&2&9&3330&3421&287&5693&301&468&206&\textbf{14400}\\
    \textbf{8}&1968&6&7&7808&11932&484&18375&642&837&575&\textbf{42634}\\
    \textbf{9}&5624&2&6&20714&48437&719&73475&1667&1483&1937&\textbf{154064}\\
    \textbf{10}&4892&3&2&38028&36810&4073&97170&13621&12970&17190&\textbf{224759}\\
    \textbf{Total}  &\textbf{13581}&\textbf{38}&\textbf{93}&\textbf{71761}&\textbf{101994}&\textbf{5794}&\textbf{197210}&\textbf{16454}&\textbf{16101}&\textbf{20045}  &\textbf{443071} \\
    \bottomrule

    \end{tabular}
\caption{\label{tab:composition}
Molecule counts in the QCDGE database categorized by element compositions and heavy atom counts.
Notably,
the numbers of molecules with specific element compositions
are 9(O), 1(F), 4(OF), 12(NF), and 9(NOF), which are
excluded from this table for clarity.
}
\end{table}

\begin{table}[ht]
    \centering

    \begin{tabularx}{\textwidth}{|X|>{\centering\arraybackslash}m{3cm}|>{\centering\arraybackslash}m{2cm}|m{1.5cm}|X|>{\centering\arraybackslash}m{3cm}|>{\centering\arraybackslash}m{2cm}|m{1.5cm}|}
    \hline
    No. & Image & Murcko Scaffold Smiles & Number of Molecules & No. & Image & Murcko Scaffold Smiles & Number of Molecules \\ \hline
    1 &\includegraphics[width=2cm,height=1.8cm,keepaspectratio]{1.eps} & C1CC1 & 17600 & 11 &\includegraphics[width=2cm,height=1.8cm,keepaspectratio]{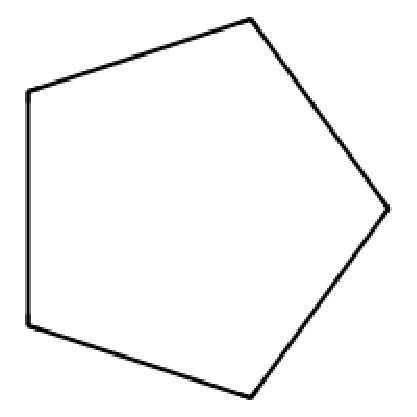} & C1CCCC1 & 3675 \\ \hline
    2 &\includegraphics[width=2cm,height=1.8cm,keepaspectratio]{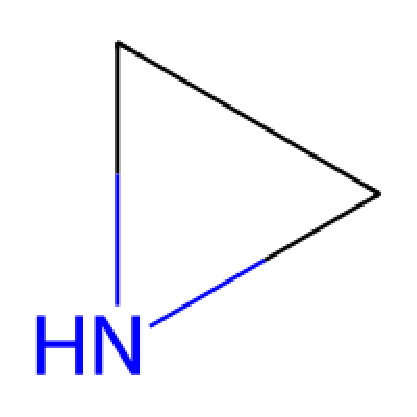} & C1CN1 & 7855 & 12 &\includegraphics[width=2cm,height=1.8cm,keepaspectratio]{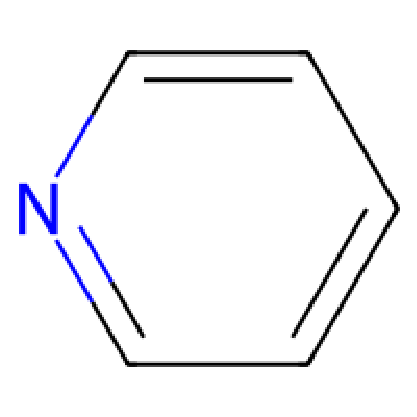} & c1ccncc1 & 3655 \\ \hline
    3 &\includegraphics[width=2cm,height=1.8cm,keepaspectratio]{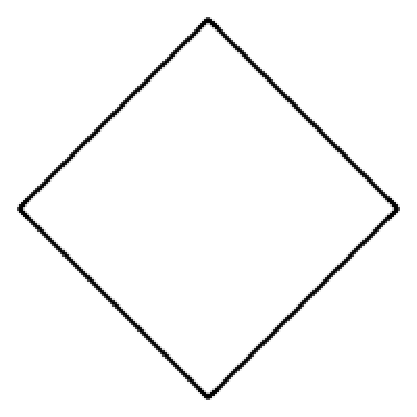} & C1CCC1 & 6814 & 13 &\includegraphics[width=2cm,height=1.8cm,keepaspectratio]{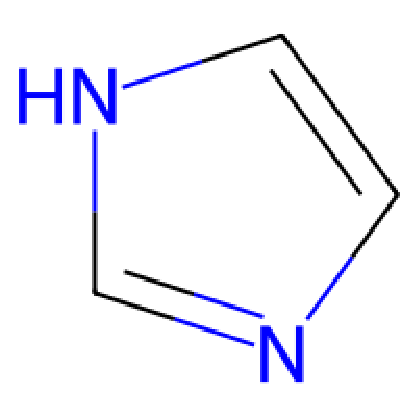} & c1c[nH]cn1 & 3541 \\ \hline
    4 &\includegraphics[width=2cm,height=1.8cm,keepaspectratio]{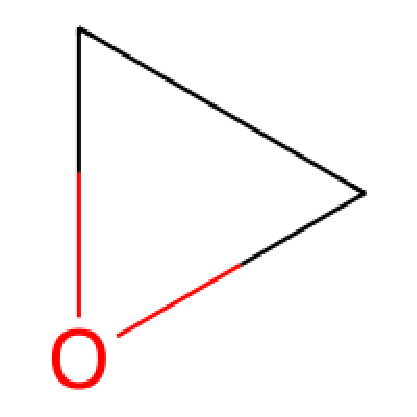} & C1CO1 & 5610 & 14 &\includegraphics[width=2cm,height=1.8cm,keepaspectratio]{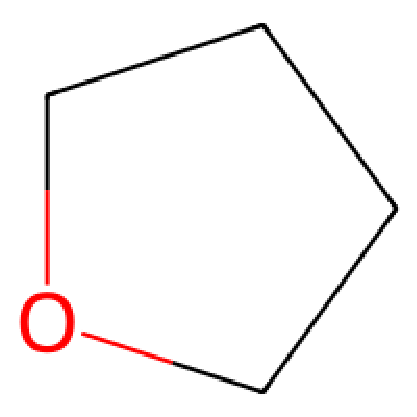} & C1CCOC1 & 3455 \\ \hline
    5 &\includegraphics[width=2cm,height=1.8cm,keepaspectratio]{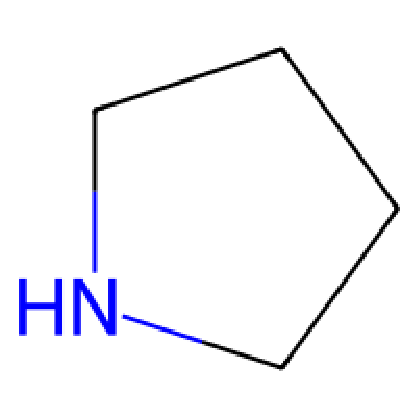} & C1CCNC1 & 5085 & 15 &\includegraphics[width=2cm,height=1.8cm,keepaspectratio]{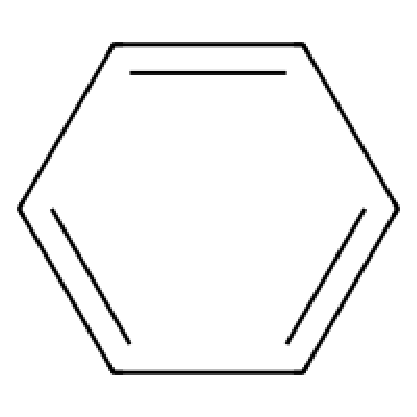} & c1ccccc1 & 3402 \\ \hline
    6 &\includegraphics[width=2cm,height=1.8cm,keepaspectratio]{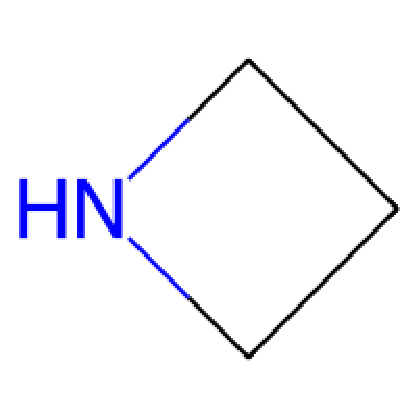} & C1CNC1 & 4633 & 16 &\includegraphics[width=2cm,height=1.8cm,keepaspectratio]{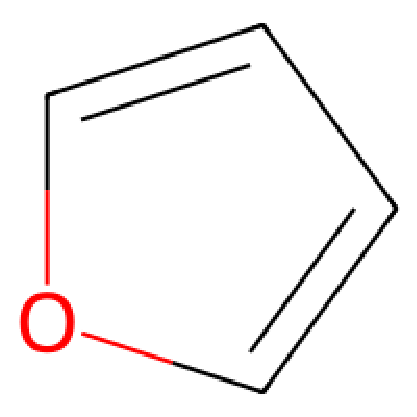} & c1ccoc1 & 3082 \\ \hline
    7 &\includegraphics[width=2cm,height=1.8cm,keepaspectratio]{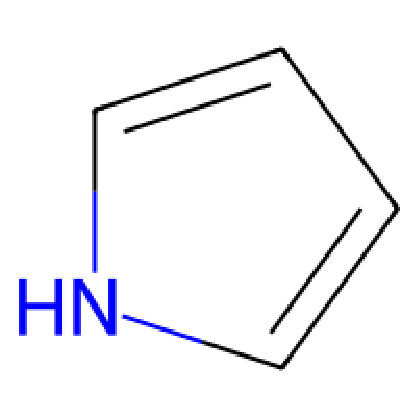} & c1cc[nH]c1 & 4596 & 17 &\includegraphics[width=2cm,height=1.8cm,keepaspectratio]{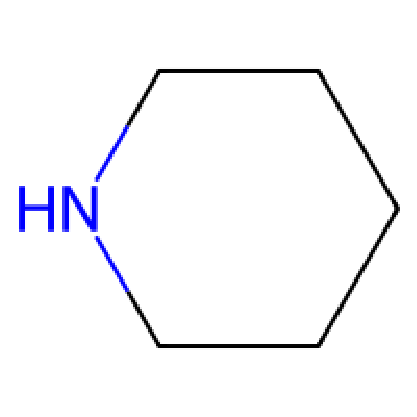} & C1CCNCC1 & 2894 \\ \hline
    8 &\includegraphics[width=2cm,height=1.8cm,keepaspectratio]{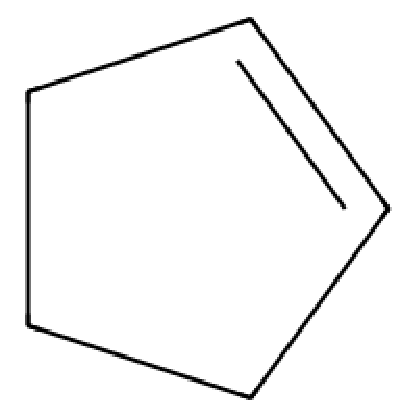} & C1=CCCC1 & 4117 & 18 &\includegraphics[width=2cm,height=1.8cm,keepaspectratio]{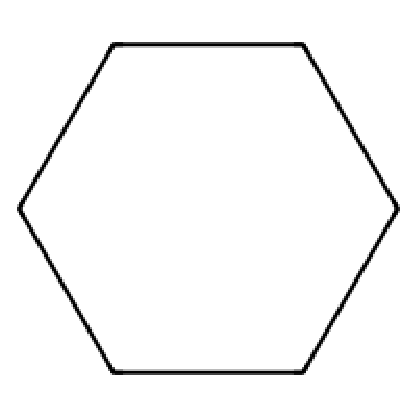} & C1CCCCC1 & 2561 \\ \hline
    9 &\includegraphics[width=2cm,height=1.8cm,keepaspectratio]{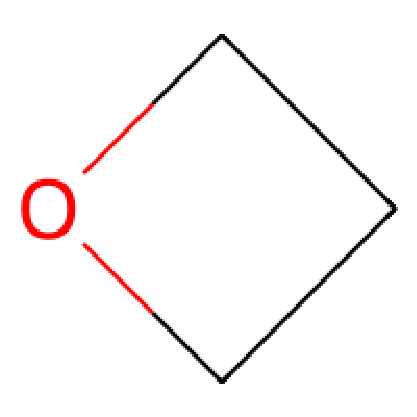} & C1COC1 & 4057 & 19 &\includegraphics[width=2cm,height=1.8cm,keepaspectratio]{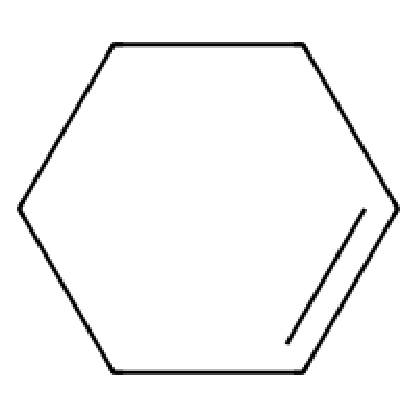} & C1=CCCCC1 & 2163 \\ \hline
    10 &\includegraphics[width=2cm,height=1.8cm,keepaspectratio]{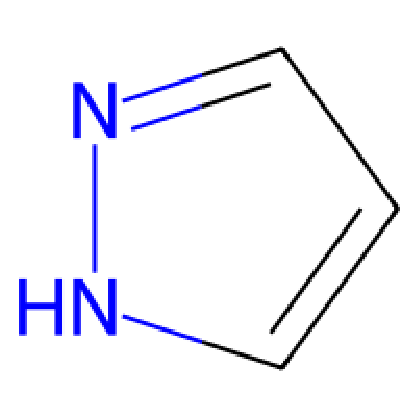} & c1cn[nH]c1 & 3885 & 20 &\includegraphics[width=2cm,height=1.8cm,keepaspectratio]{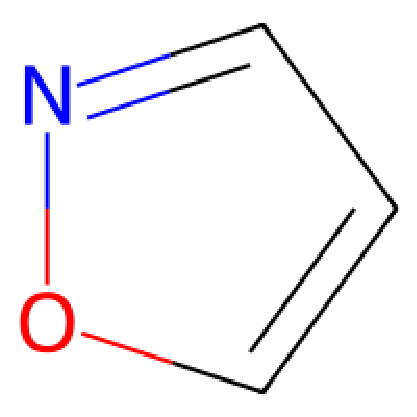} & c1cnoc1 & 2068 \\ \hline

    \end{tabularx}
    \caption{
        Top 20 Murcko scaffold SMILES in QCDGE database, along with their corresponding images and quantities.
    }
    \label{tab:all_atom}
\end{table}

\begin{table}[ht]
    \centering

    \begin{tabularx}{\textwidth}{|X|>{\centering\arraybackslash}m{2cm}|>{\centering\arraybackslash}m{2.6cm}|m{1.3cm}|X|>{\centering\arraybackslash}m{2cm}|>{\centering\arraybackslash}m{2.6cm}|m{1.3cm}|}
    \hline
    No. & Image & Murcko Scaffold Smiles & Number of Molecules & No. & Image & Murcko Scaffold Smiles & Number of Molecules \\ \hline
    1 &\includegraphics[width=2cm,height=1.7cm,keepaspectratio]{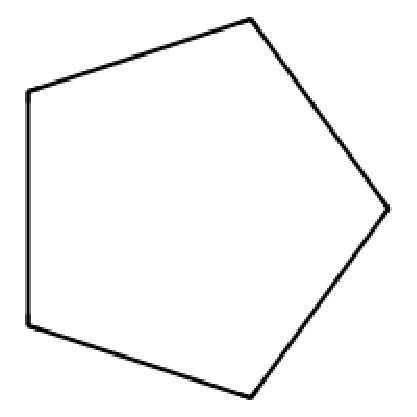} & C1CCCC1 & 53327 & 11 &\includegraphics[width=2cm,height=1.7cm,keepaspectratio]{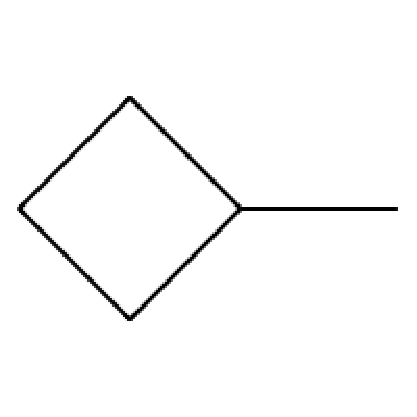} & CC1CCC1 & 4960 \\ \hline
    2 &\includegraphics[width=2cm,height=1.7cm,keepaspectratio]{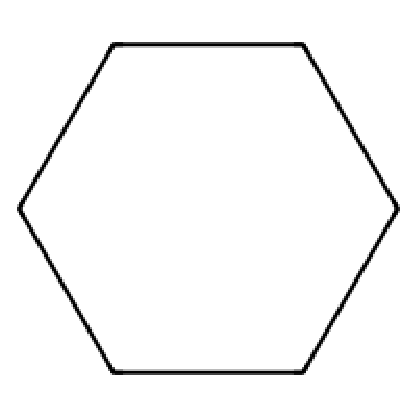} & C1CCCCC1 & 33802 & 12 &\includegraphics[width=2cm,height=1.7cm,keepaspectratio]{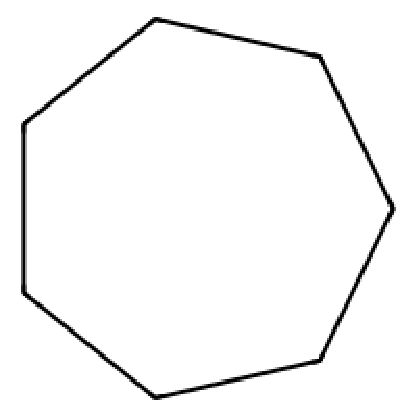} & C1CCCCCC1 & 4902 \\ \hline
    3 &\includegraphics[width=2cm,height=1.7cm,keepaspectratio]{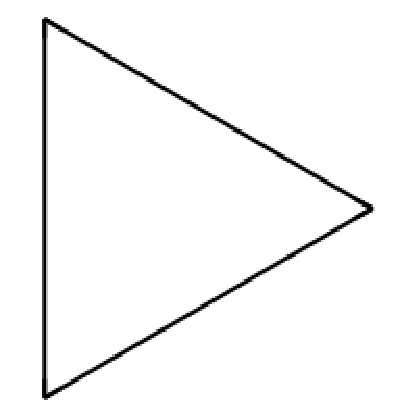} & C1CC1 & 31767 & 13 &\includegraphics[width=2cm,height=1.7cm,keepaspectratio]{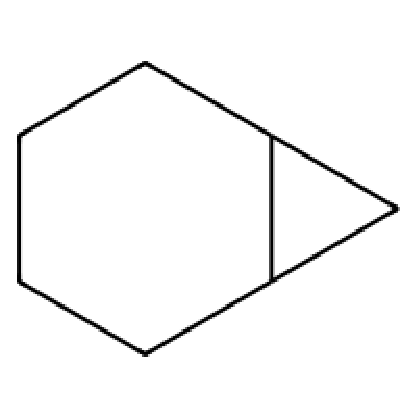} & C1CCC2CC2C1 & 4001 \\ \hline
    4 &\includegraphics[width=2cm,height=1.7cm,keepaspectratio]{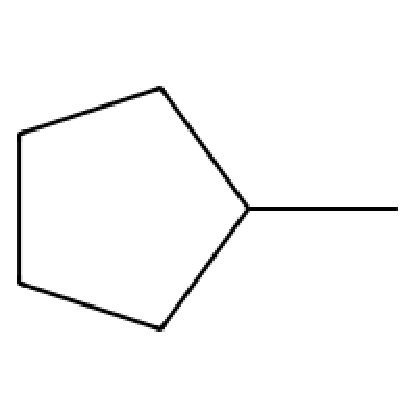} & CC1CCCC1 & 17570 & 14 &\includegraphics[width=2cm,height=1.7cm,keepaspectratio]{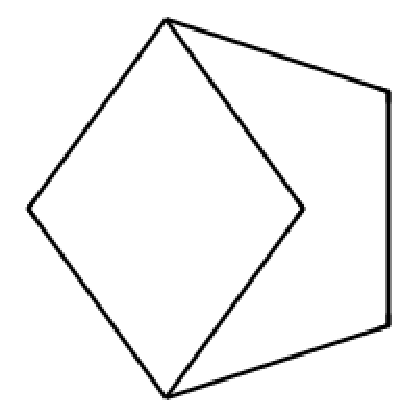} & C1CC2CC1C2 & 3524 \\ \hline
    5 &\includegraphics[width=2cm,height=1.7cm,keepaspectratio]{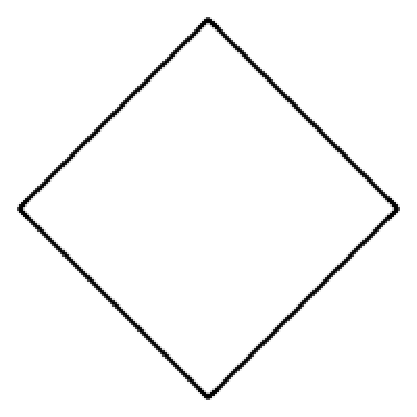} & C1CCC1 & 16435 & 15 &\includegraphics[width=2cm,height=1.7cm,keepaspectratio]{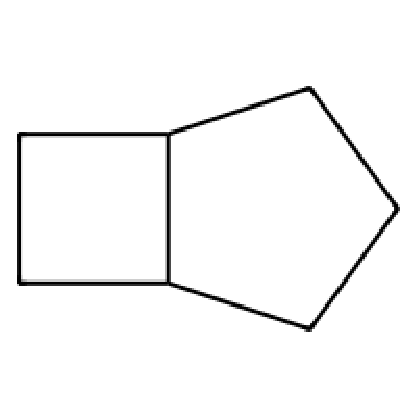} & C1CC2CCC2C1 & 3455 \\ \hline
    6 &\includegraphics[width=2cm,height=1.7cm,keepaspectratio]{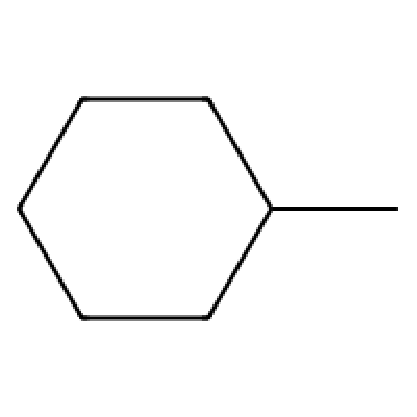} & CC1CCCCC1 & 12189 & 16 &\includegraphics[width=2cm,height=1.7cm,keepaspectratio]{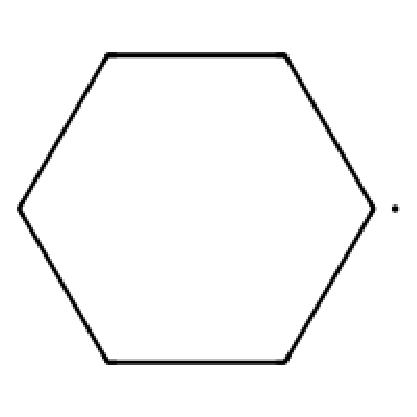} & [CH]1CCCCC1 & 2933 \\ \hline
    7 &\includegraphics[width=2cm,height=1.7cm,keepaspectratio]{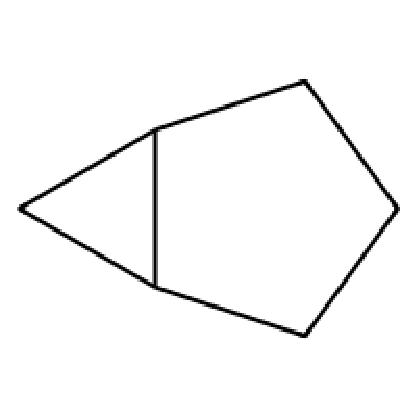} & C1CC2CC2C1 & 8816 & 17 &\includegraphics[width=2cm,height=1.7cm,keepaspectratio]{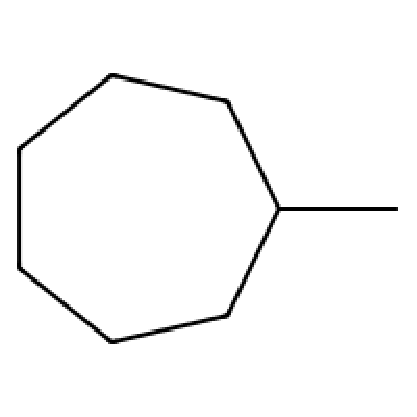} & CC1CCCCCC1 & 2851 \\ \hline
    8 &\includegraphics[width=2cm,height=1.7cm,keepaspectratio]{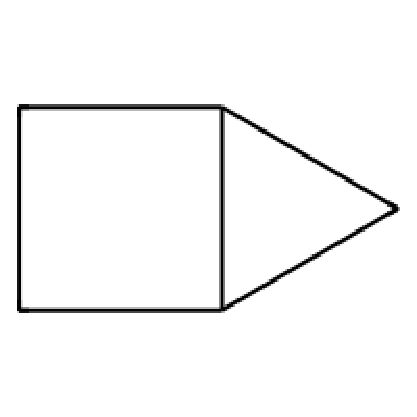} & C1CC2CC12 & 5886 & 18 &\includegraphics[width=2cm,height=1.7cm,keepaspectratio]{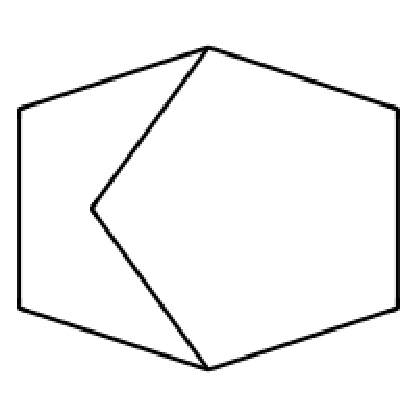} & C1CC2CCC1C2 & 2778 \\ \hline
    9 &\includegraphics[width=2cm,height=1.7cm,keepaspectratio]{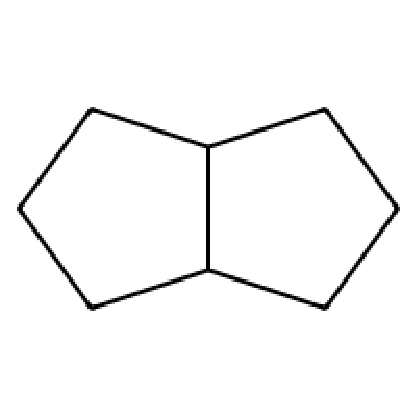} & C1CC2CCCC2C1 & 5139 & 19 &\includegraphics[width=2cm,height=1.7cm,keepaspectratio]{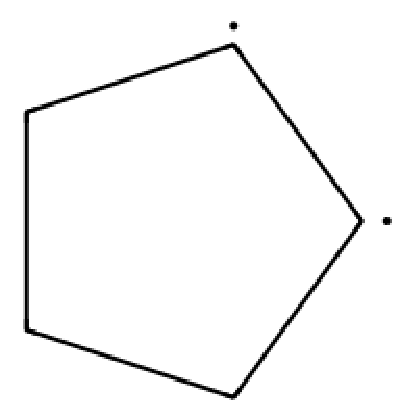} & [CH]1[CH]CCC1 & 2642 \\ \hline
    10 &\includegraphics[width=2cm,height=1.7cm,keepaspectratio]{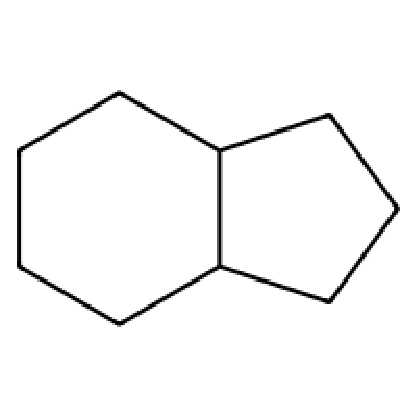} & C1CCC2CCCC2C1 & 5110 & 20 &\includegraphics[width=2cm,height=1.7cm,keepaspectratio]{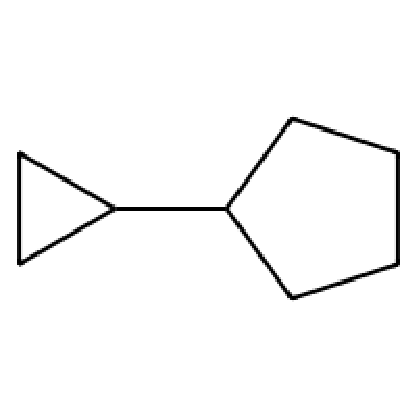} & C1CCC(C2CC2)C1 & 2413 \\ \hline

    \end{tabularx}
    \caption{\label{tab:all_C}
        Make Murcko scaffolds generic, where all atom types are transformed into carbon (C) and all bonds are considered as single bonds.
        Top 20 generic Murcko scaffold SMILES in the QCDGE database, along with their corresponding images and quantities.
    }
\end{table}

\begin{table}[ht]
    \centering

    \begin{tabularx}{\textwidth}{|X|>{\centering\arraybackslash}m{3.2cm}|>{\centering\arraybackslash}m{4cm}|X|>{\centering\arraybackslash}m{3.2cm}|>{\centering\arraybackslash}m{4cm}|}
    \hline
    No. & General structure & Substituents & No. & General structure & Substituents \\ \hline
    1 &\includegraphics[width=2cm,height=1.8cm,keepaspectratio]{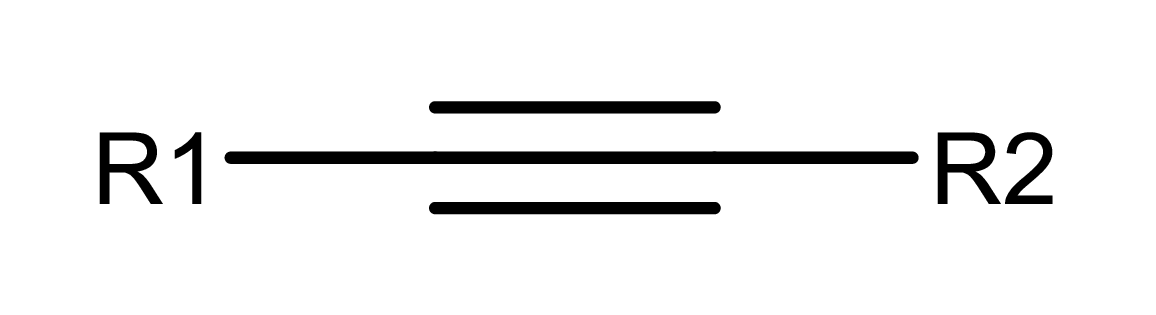} & R1, R2= H, alkyl, aryl& 11 &\includegraphics[width=2cm,height=1.8cm,keepaspectratio]{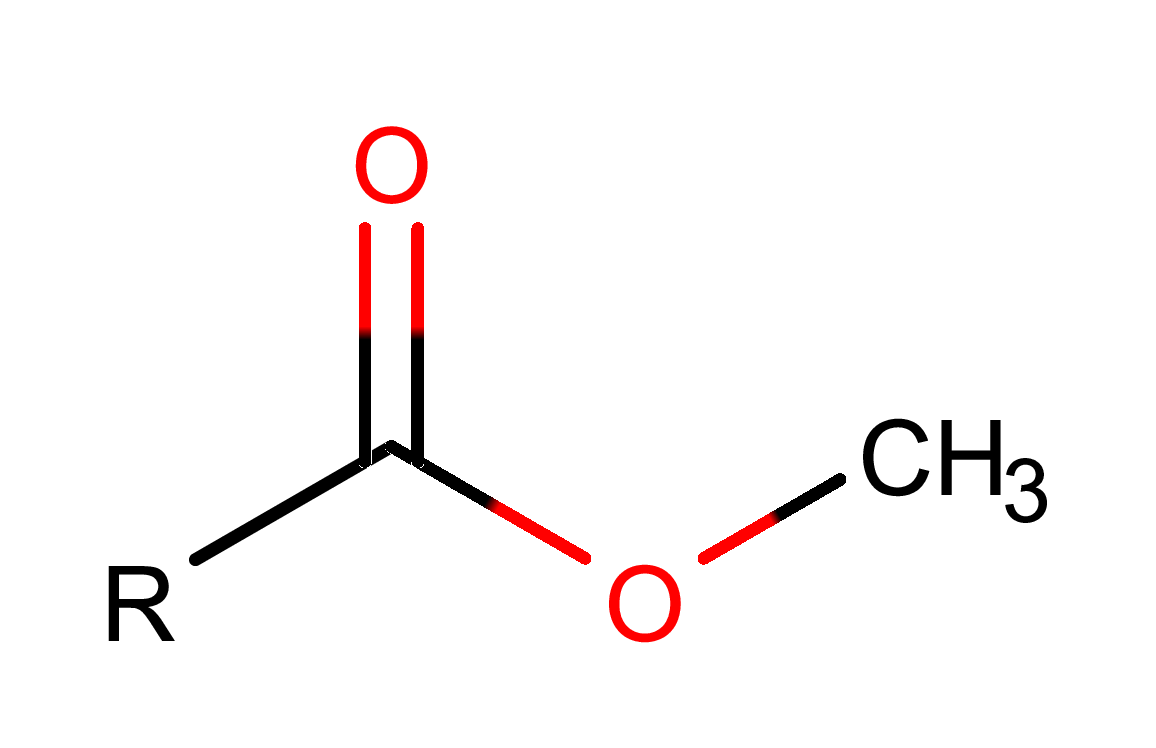} & R = H, acyl, alkyl, aryl \\ \hline
    2 &\includegraphics[width=2cm,height=1.8cm,keepaspectratio]{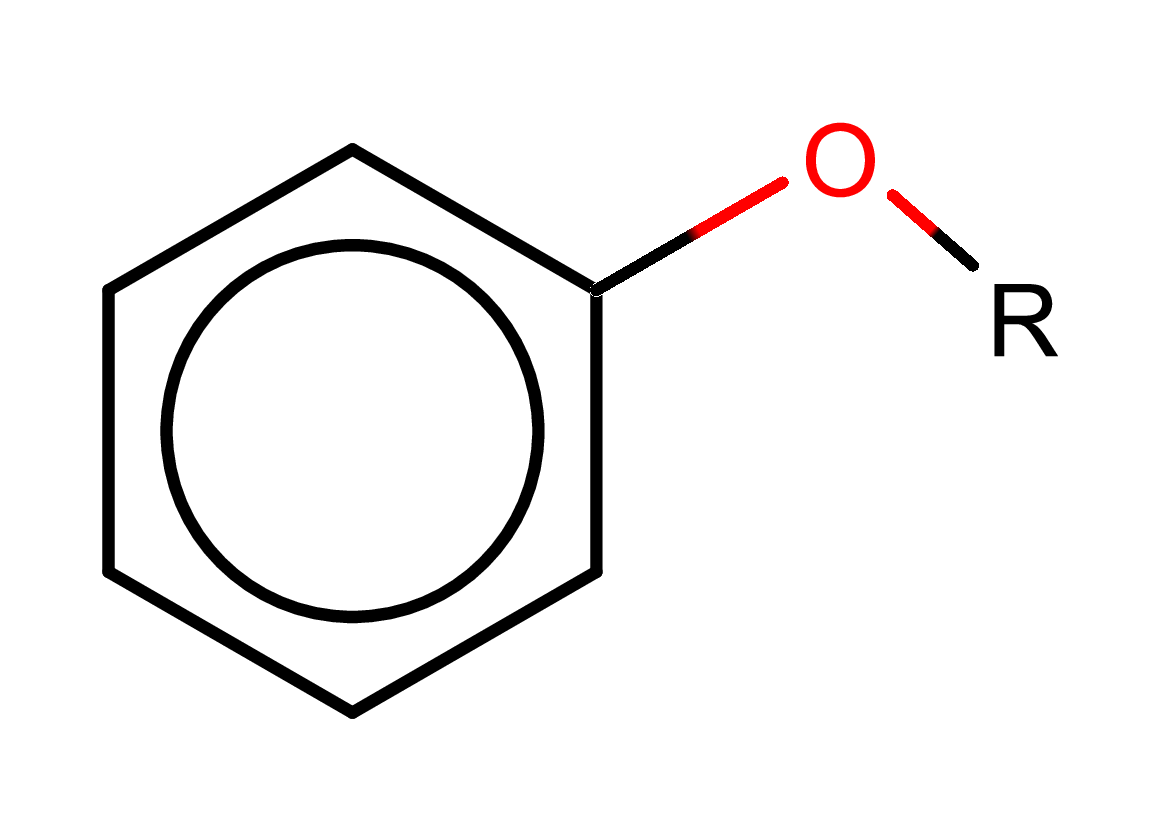} & R = H, alkyl, aryl & 12 &\includegraphics[width=2cm,height=1.8cm,keepaspectratio]{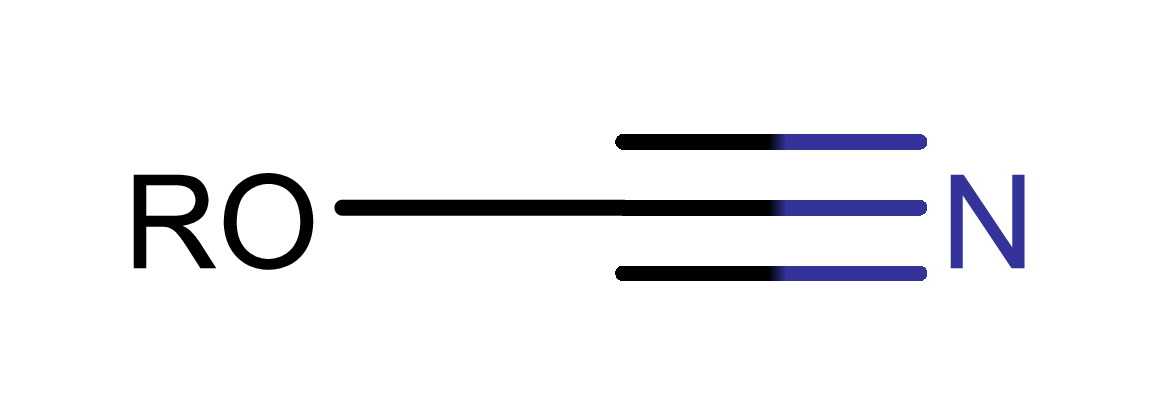} & R = H, alkyl, aryl \\ \hline
    3 &\includegraphics[width=2cm,height=1.8cm,keepaspectratio]{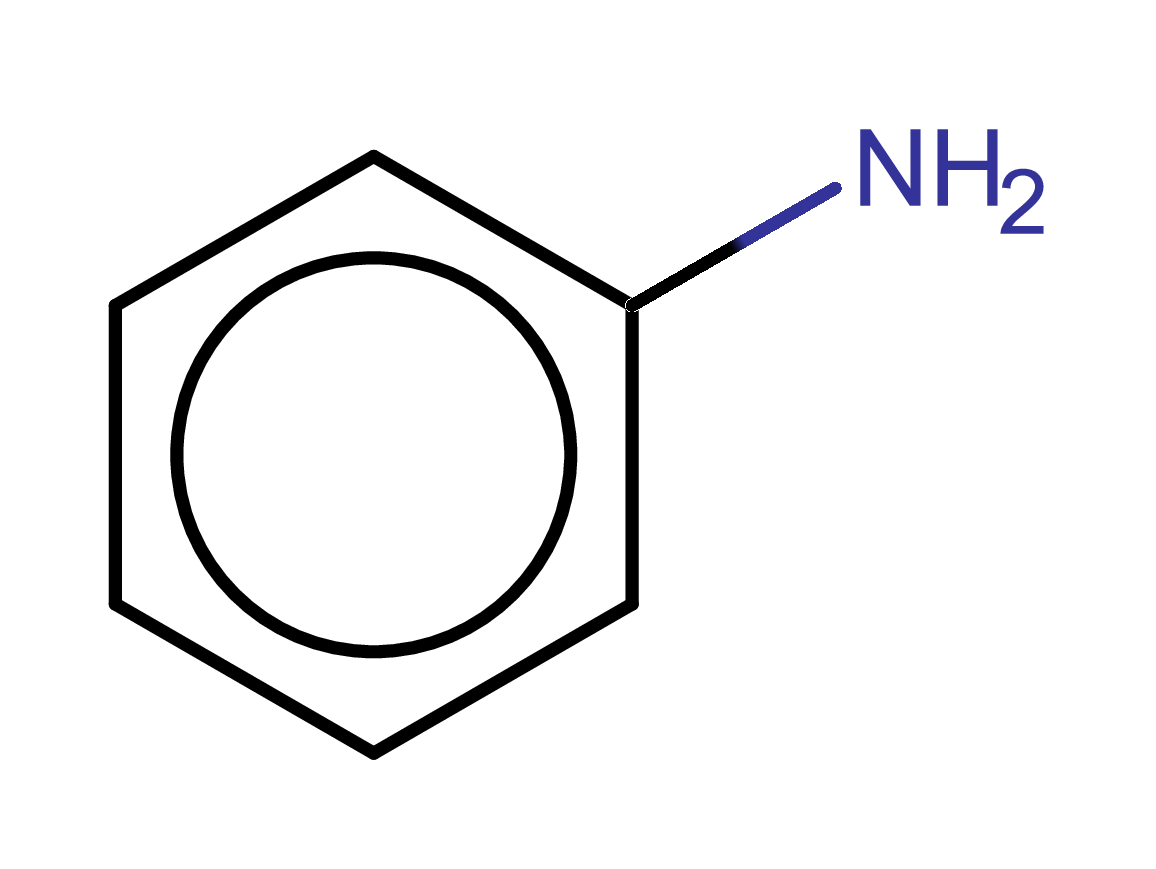} & R = aryl & 13 &\includegraphics[width=2cm,height=1.8cm,keepaspectratio]{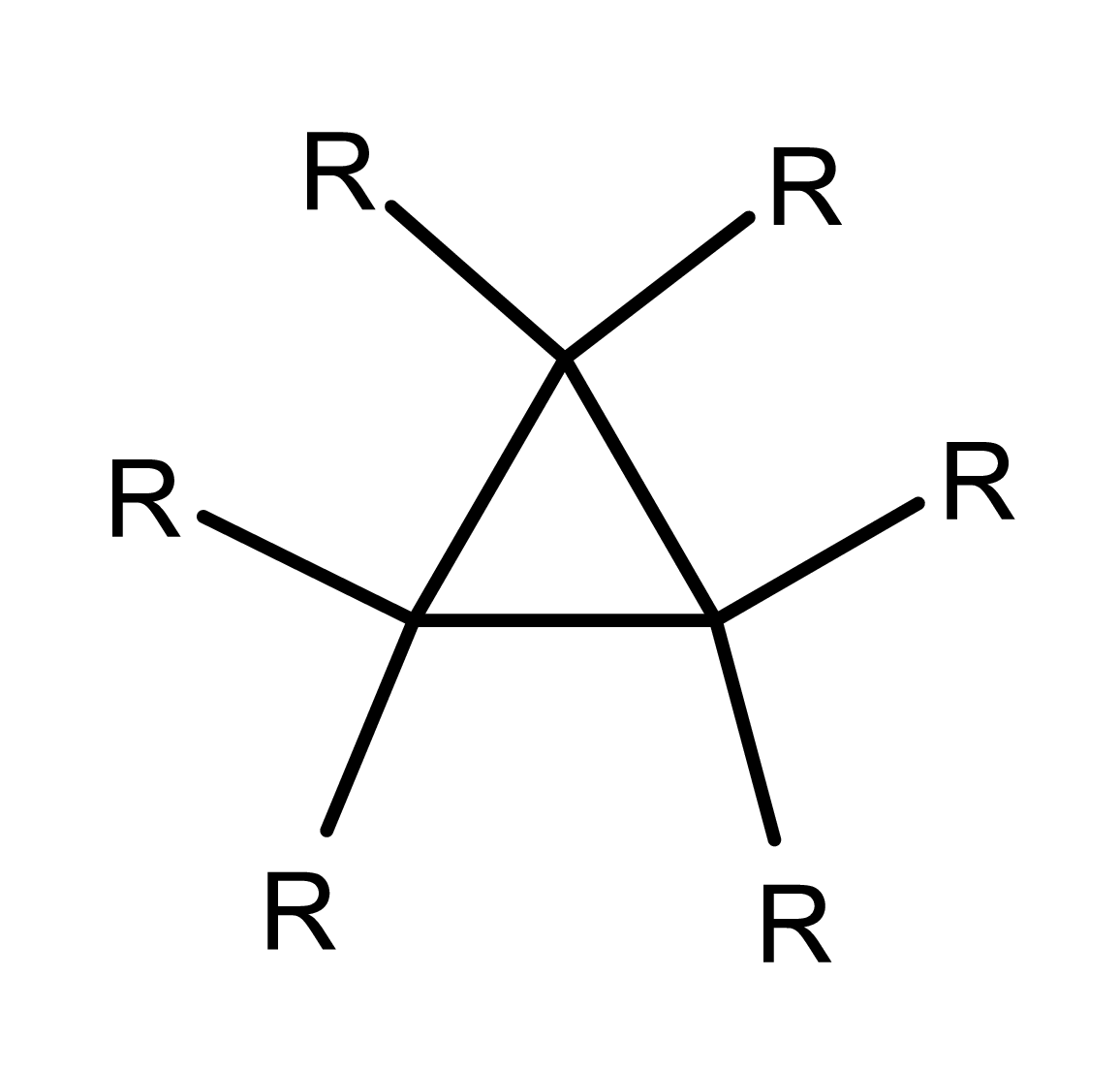} & any compound with a cyclopropyl structure \\ \hline
    4 &\includegraphics[width=2cm,height=1.8cm,keepaspectratio]{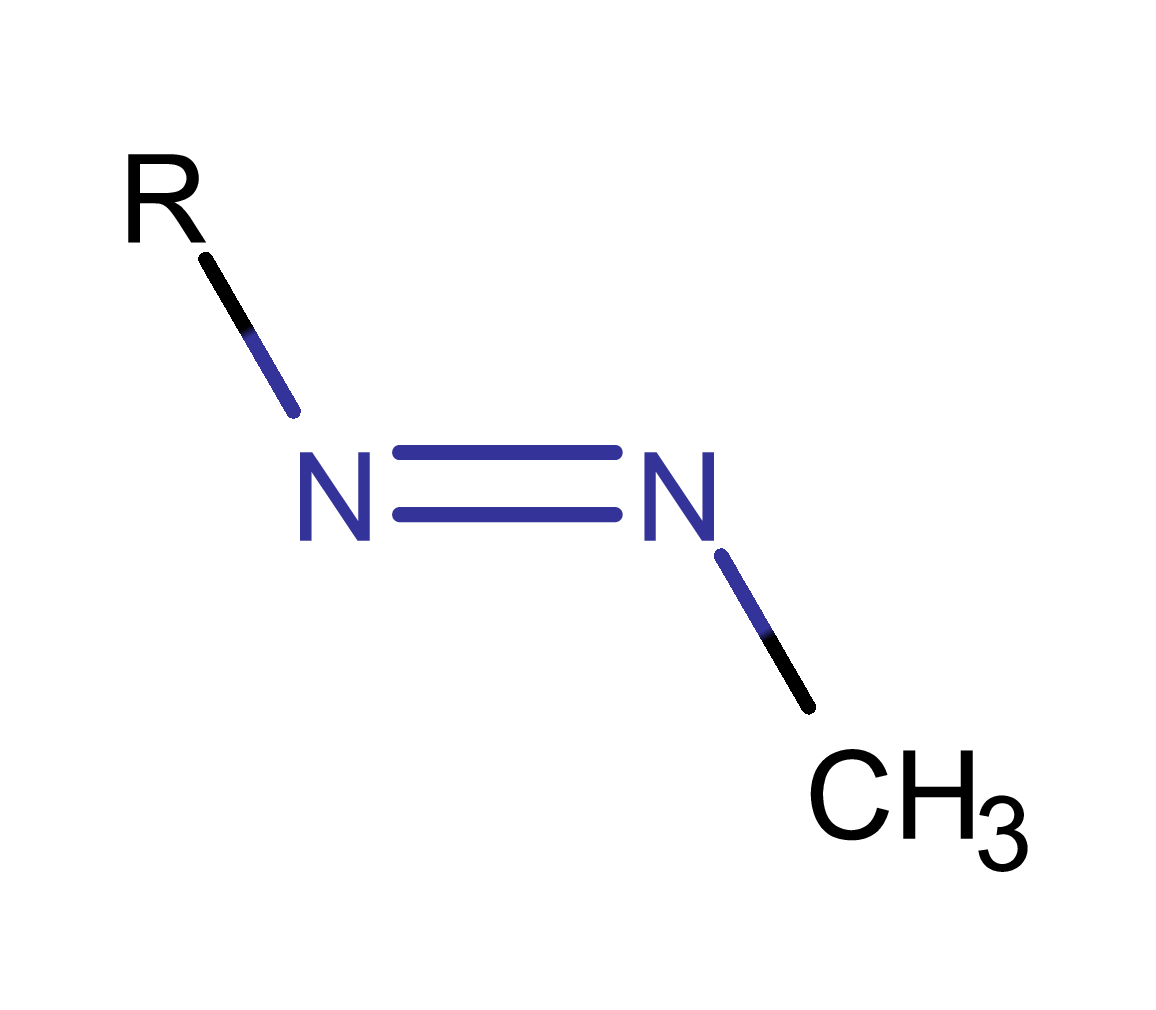} &R = H, alkyl, aryl & 14 &\includegraphics[width=2cm,height=1.8cm,keepaspectratio]{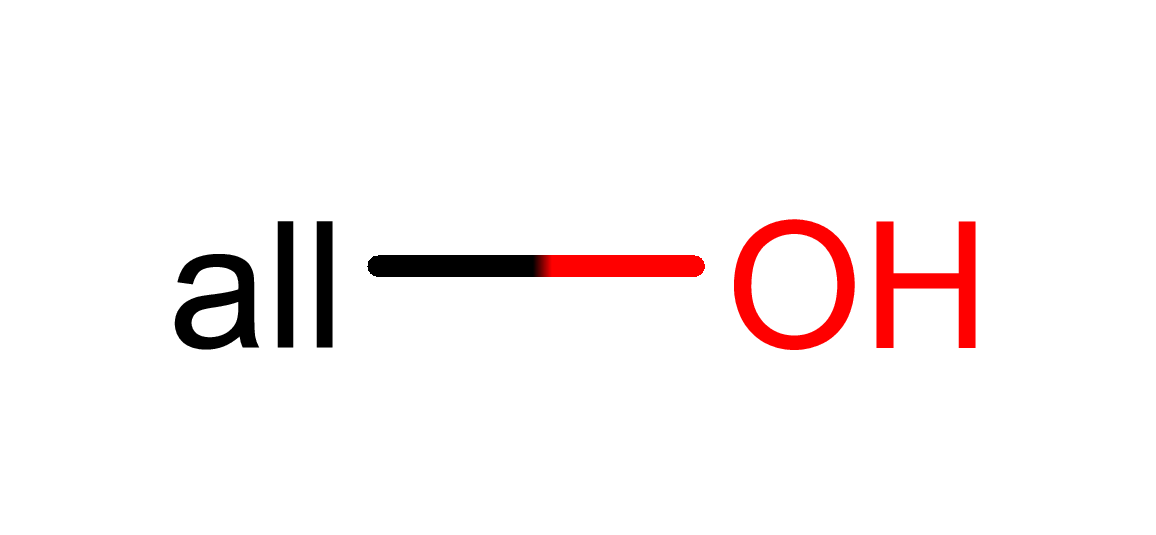} & any compound with a OH structure\\ \hline
    5 &\includegraphics[width=2cm,height=1.8cm,keepaspectratio]{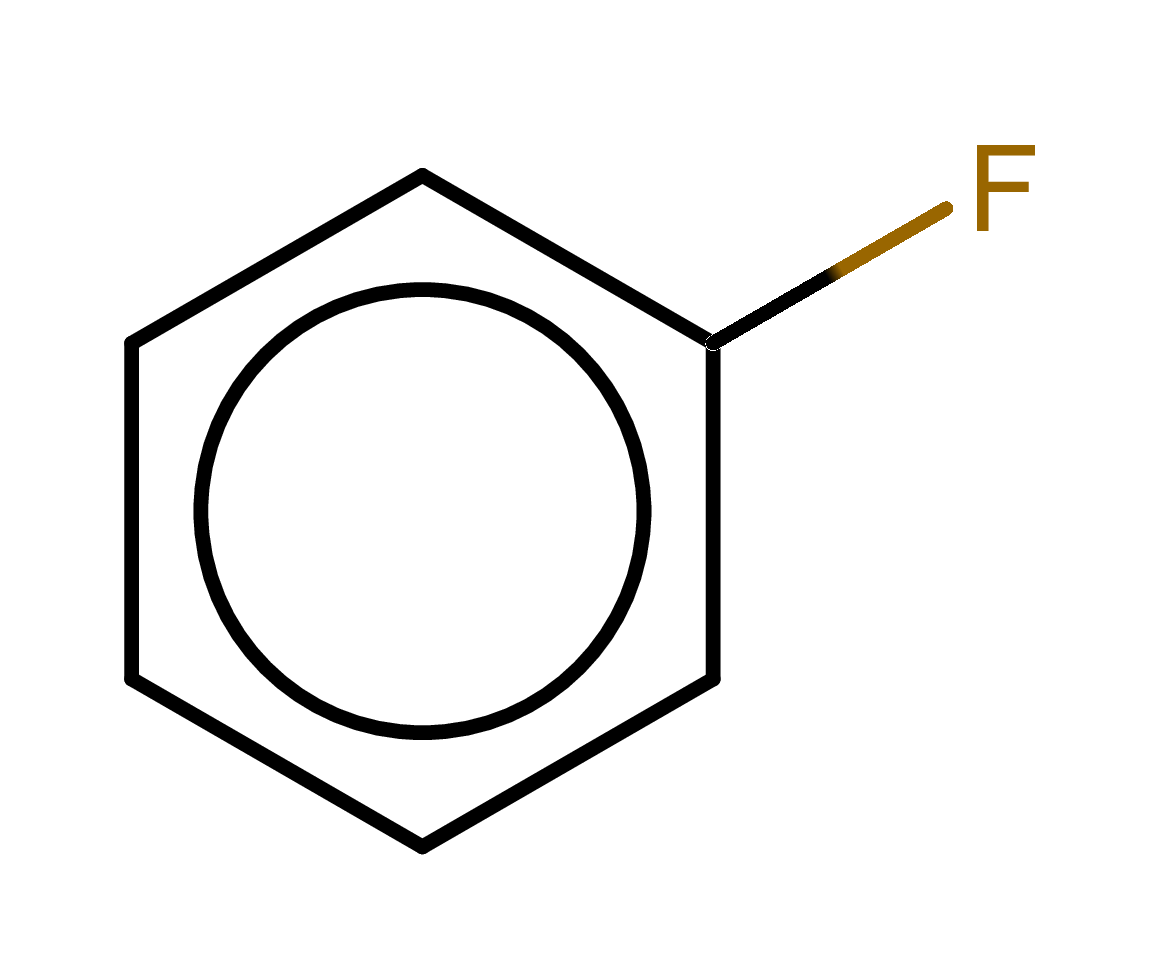} & any (hetero) aromatic compound with an F atom & 15 &\includegraphics[width=2cm,height=1.8cm,keepaspectratio]{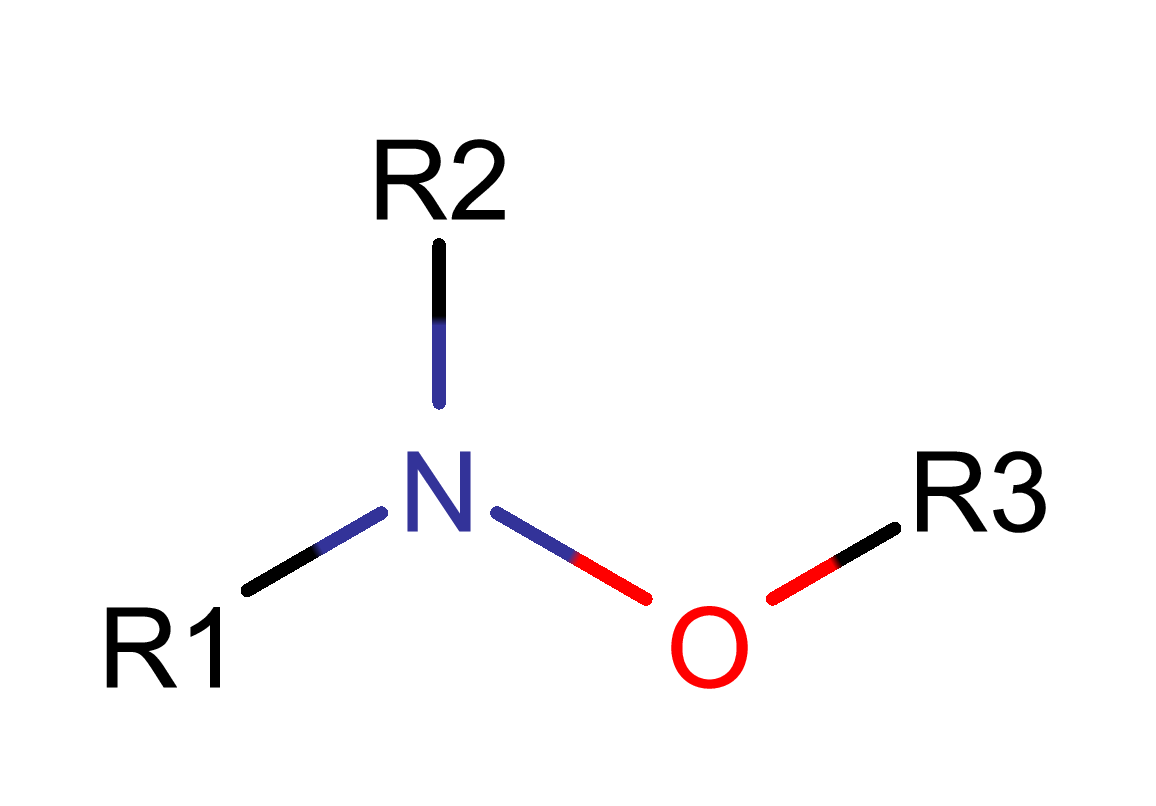} & R1, R2, R3 = H, alkyl, aryl \\ \hline
    6 &\includegraphics[width=2cm,height=1.8cm,keepaspectratio]{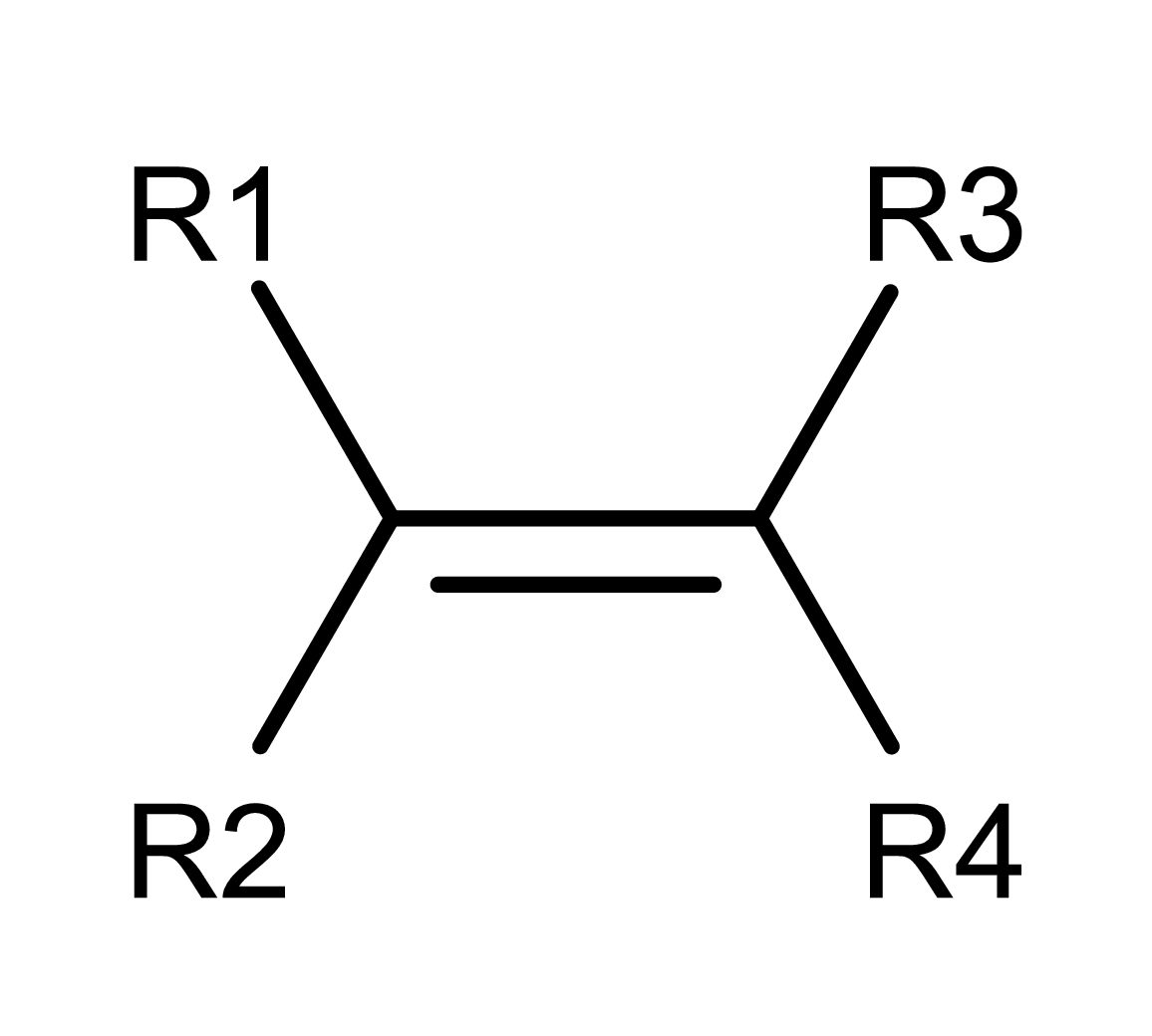} & any compound with a double bond & 16 &\includegraphics[width=2cm,height=1.8cm,keepaspectratio]{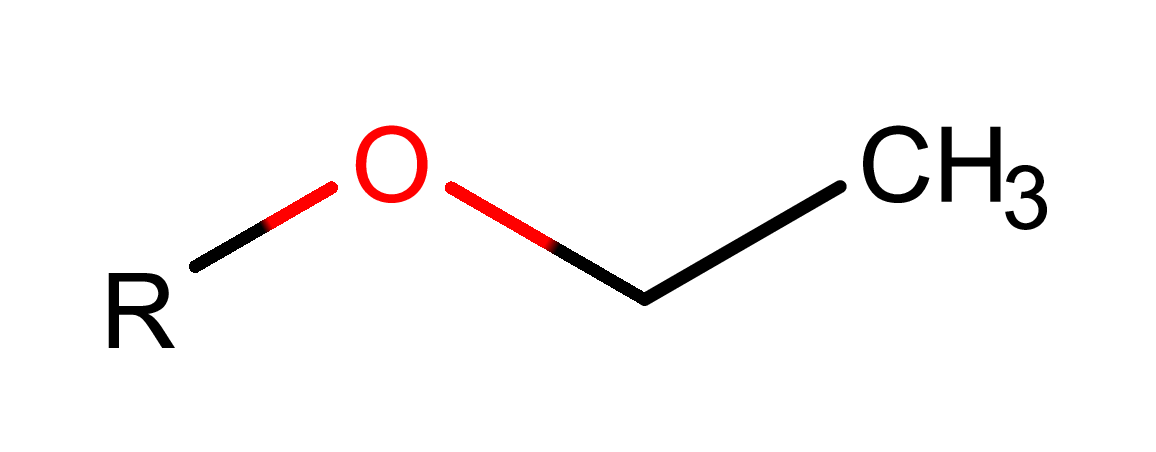} & R = alkyl, aryl \\ \hline
    7 &\includegraphics[width=2cm,height=1.8cm,keepaspectratio]{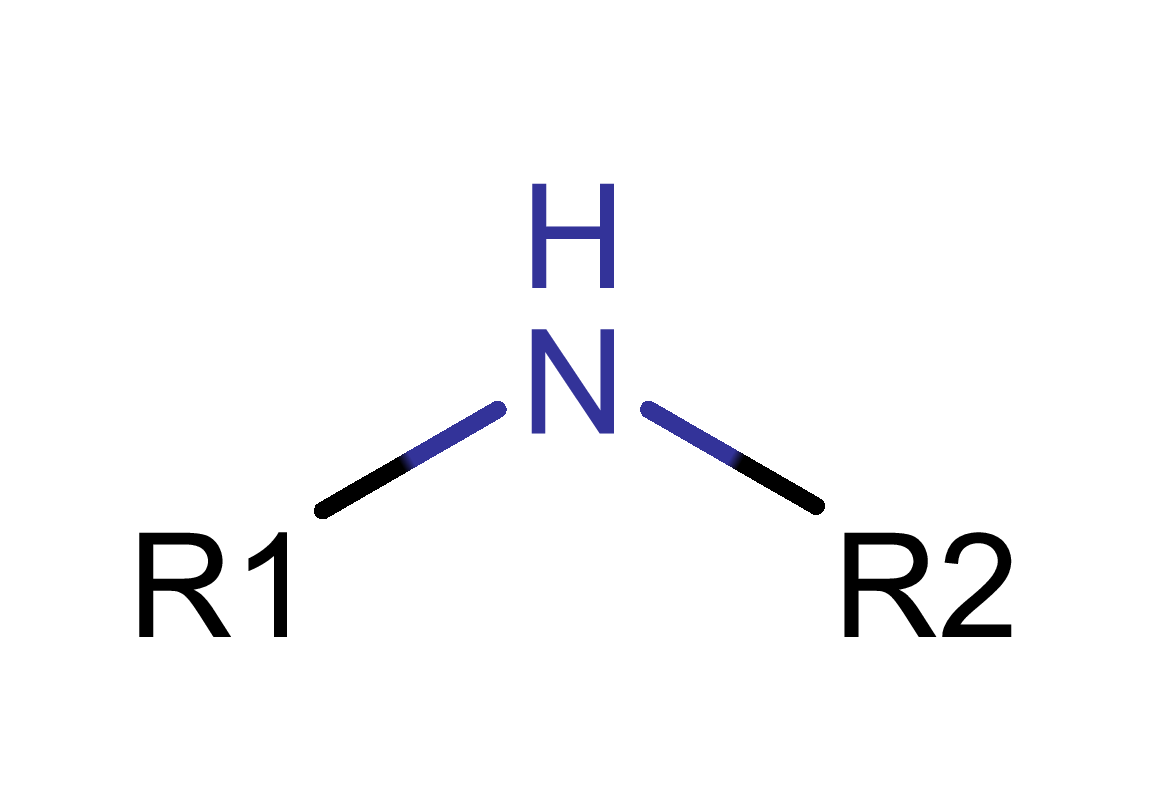} & R1, R2 = aryl & 17 &\includegraphics[width=2cm,height=1.8cm,keepaspectratio]{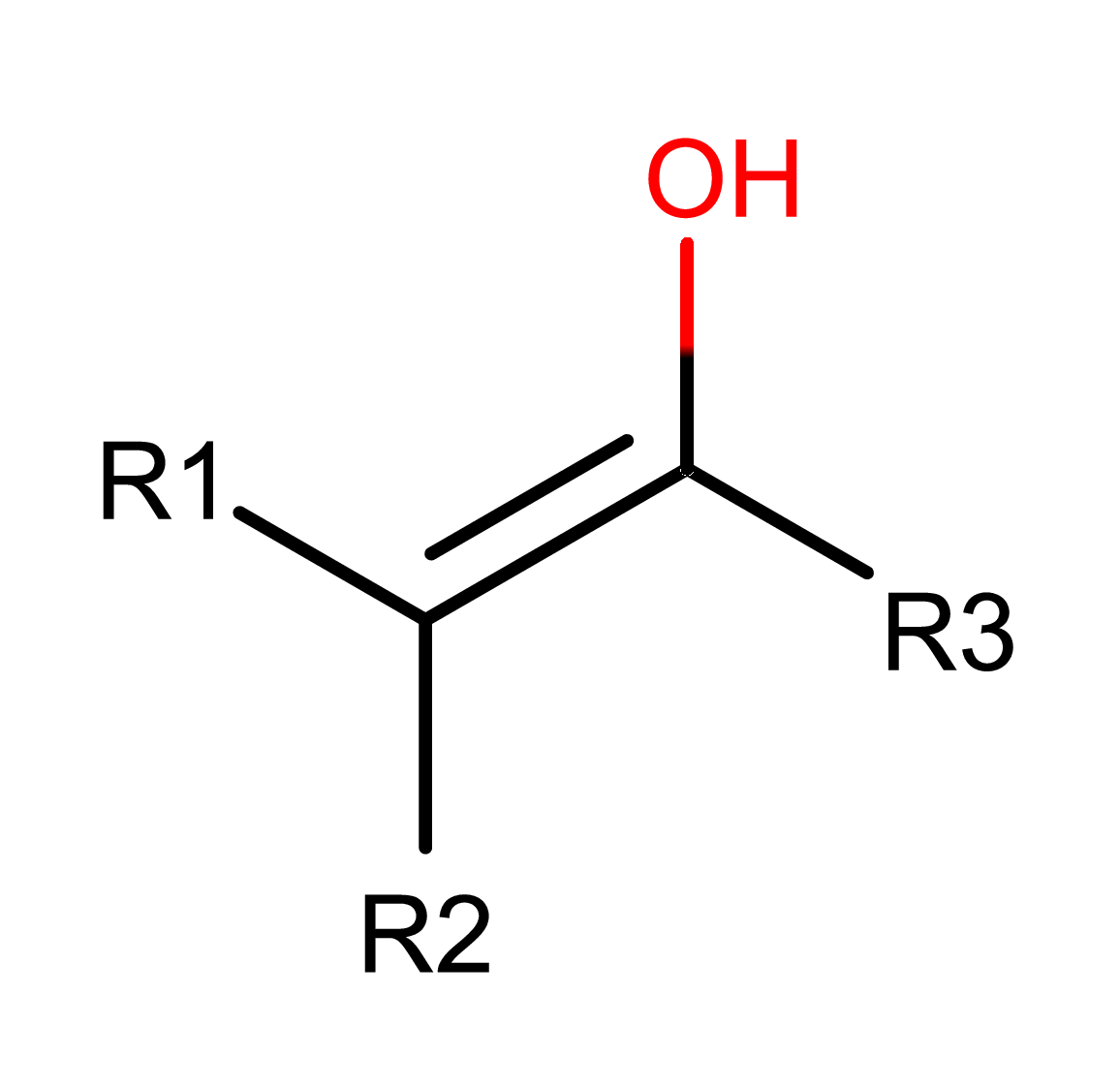} & R1, R2, R3, R4 = H, acyl, alkyl, aryl \\ \hline
    8 &\includegraphics[width=2cm,height=1.8cm,keepaspectratio]{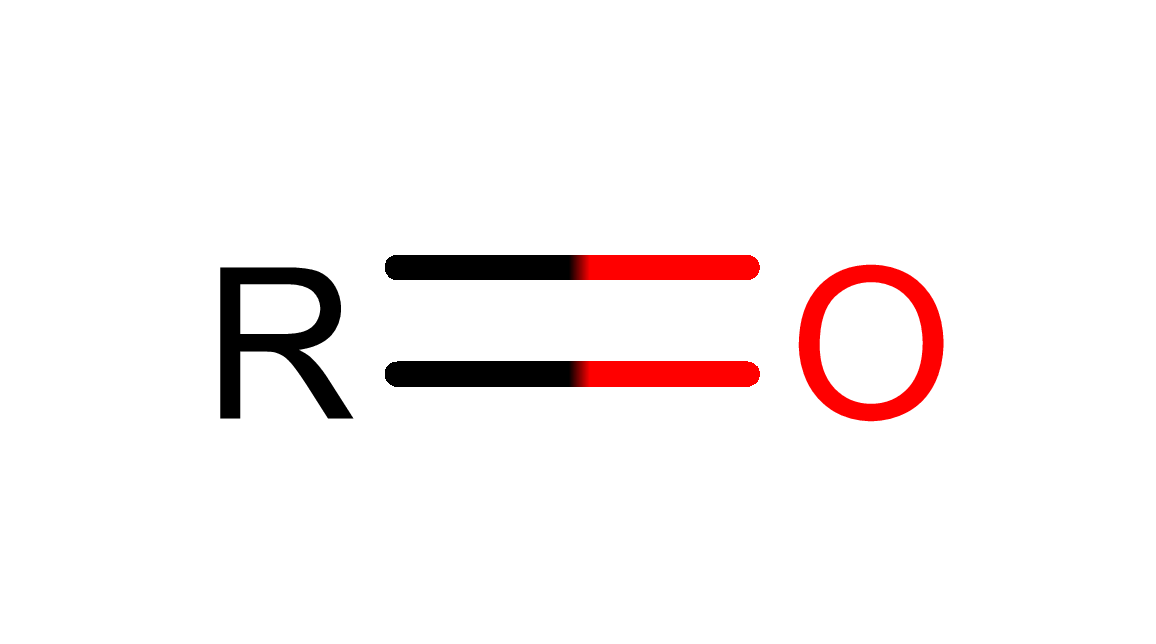} & R = H, alkyl, aryl & 18 &\includegraphics[width=2cm,height=1.8cm,keepaspectratio]{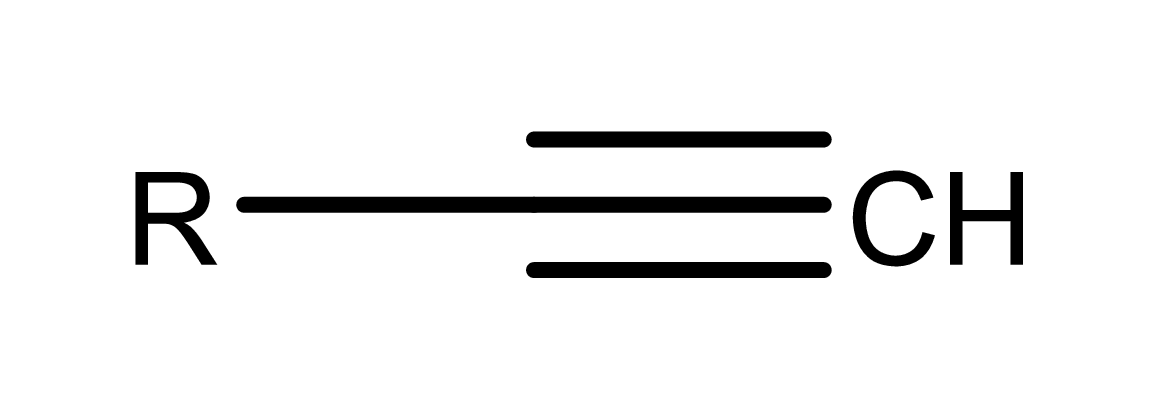} &  any compound with a triple bond \\ \hline
    9 &\includegraphics[width=2cm,height=1.8cm,keepaspectratio]{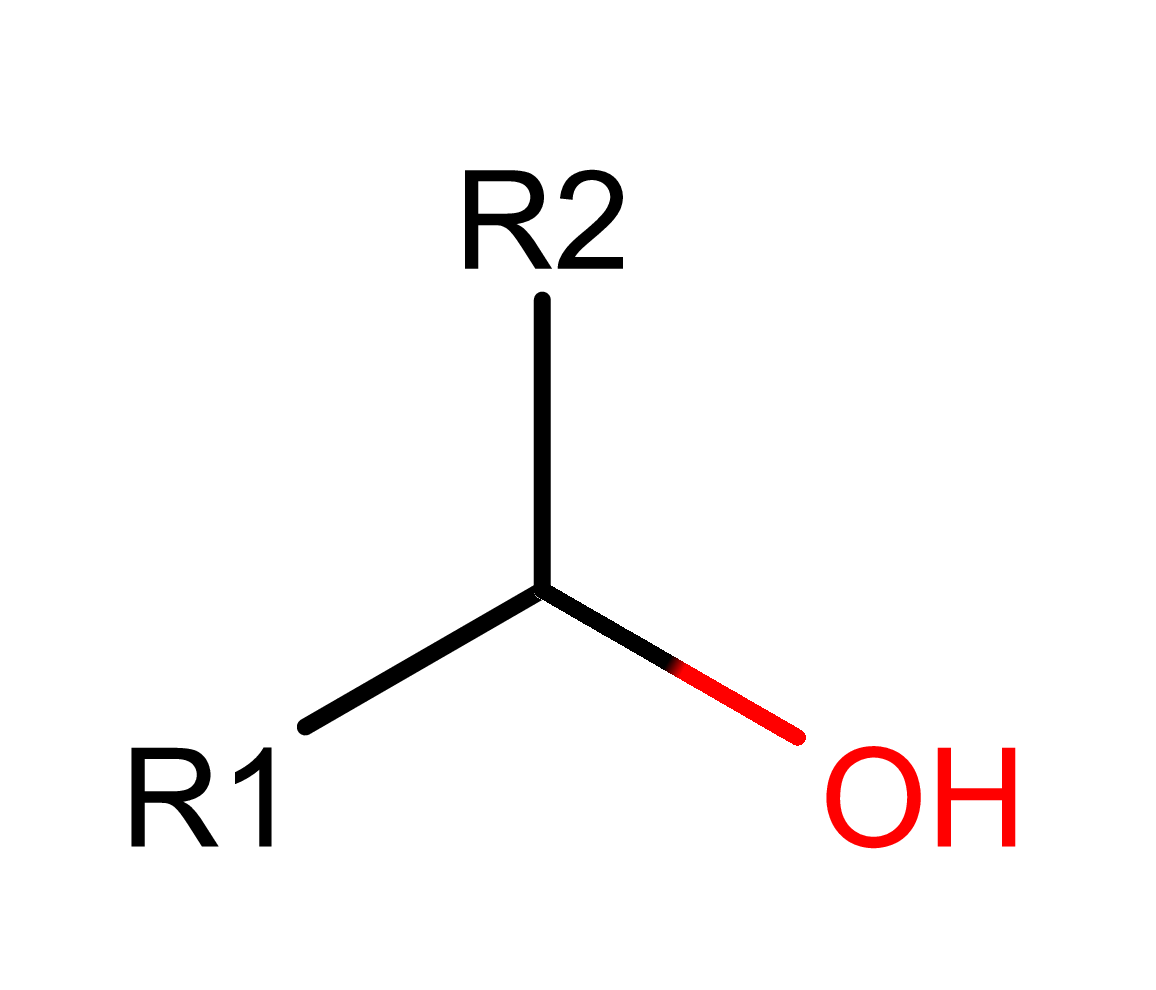} &R1, R2= alkyl, aryl & 19 &\includegraphics[width=2cm,height=1.8cm,keepaspectratio]{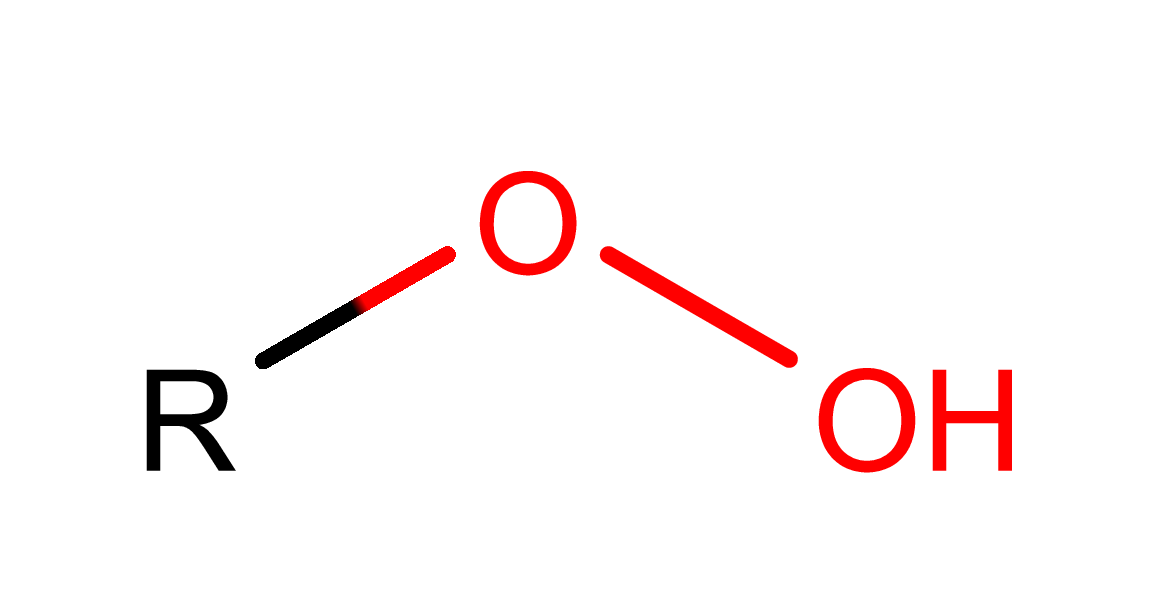} & R = H, alkyl, aryl\\ \hline
    10 &\includegraphics[width=2cm,height=1.8cm,keepaspectratio]{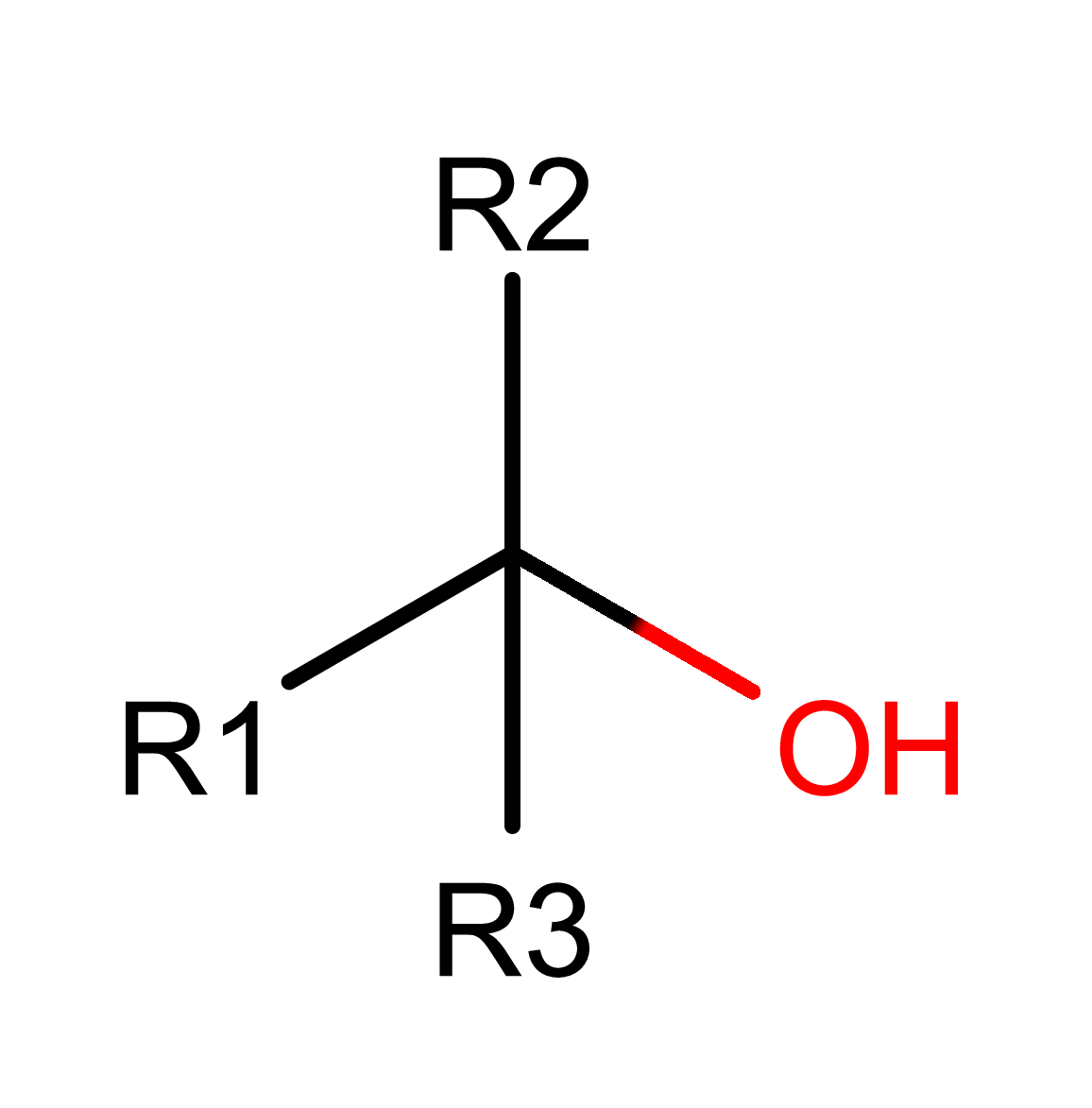} & R1, R2, R3 = H, alkyl, ary & 20 &\includegraphics[width=2cm,height=1.8cm,keepaspectratio]{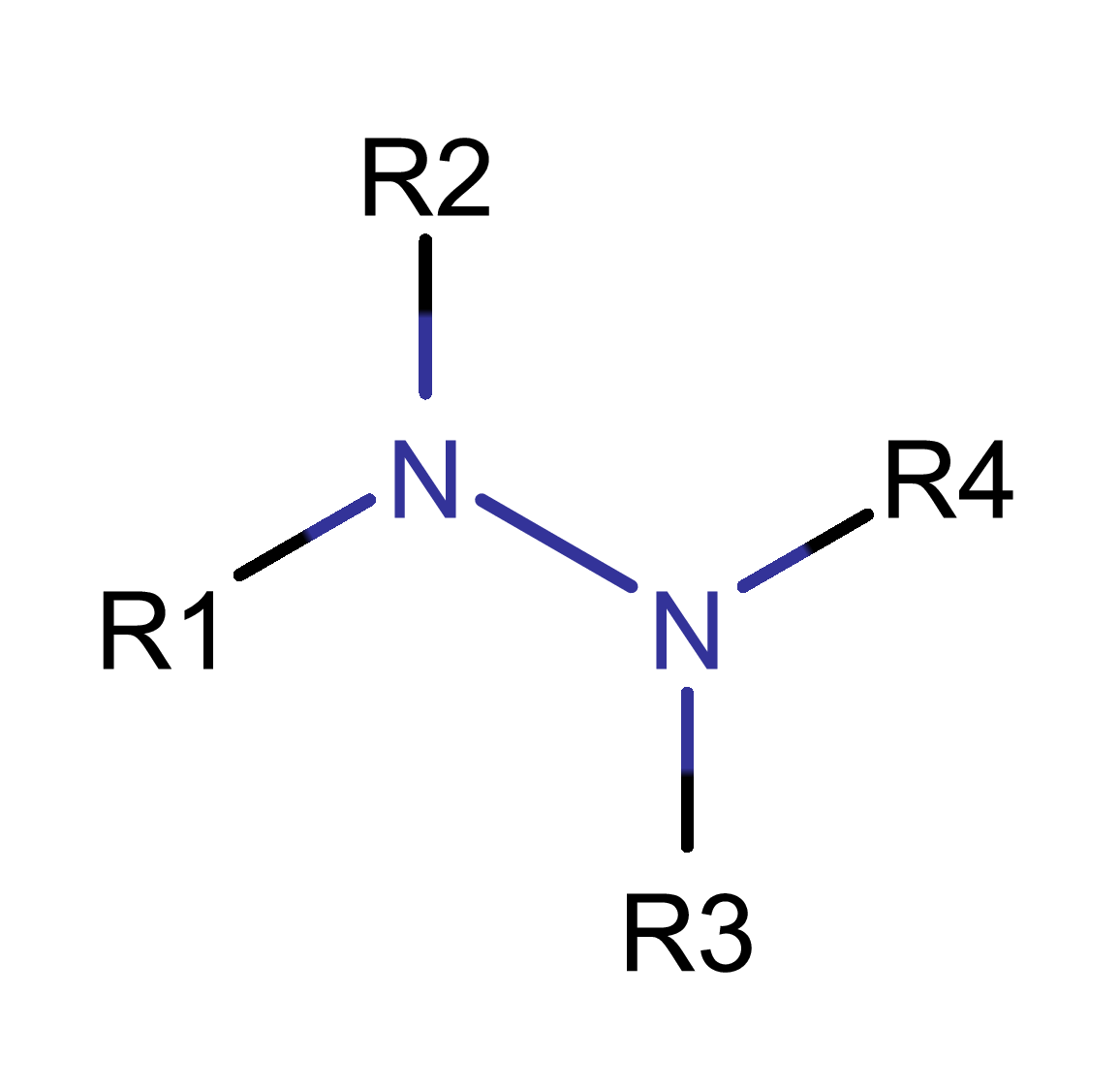} & R1, R2, R3, R4 = alkyl, aryl \\ \hline
    \end{tabularx}
    \caption{\label{tab:fgs}
        General functional structures and substitute moieties  of top 20  functional groups in the QCDGE database. Marvin \cite{marvin2024} was used for drawing general structures.
    }
\end{table}
\end{document}